\documentclass[longauth]{aa} 
\usepackage{graphicx}
\usepackage{txfonts}
\usepackage{natbib}
\usepackage{pdflscape}

\newcommand{\mygi}{MyGIsFOS}
\newcommand{\Teff}{\ensuremath{T_\mathrm{eff}}}

\newcommand{\logg}{\ensuremath{\log\,g}}
\def\teff{$T\rm_{eff}$}
\newcommand{\kms}{$\rm km s ^{-1}$}
\usepackage{hyperref}
\hypersetup{colorlinks=true,linkcolor=blue,citecolor=blue,filecolor=blue,urlcolor=blue}
\begin{document} 
\title{MINCE I. Presentation of the project and \\
of the first year sample\thanks{Based on observations made with HARPS-N at TNG, FIES at NOT, Sophie at OHP and ESPaDOnS at CFHT.},\thanks{Tables B.1, C.1-C.3  are only available at the CDS via anonymous ftp to cdsarc.u-strasbg.fr (130.79.128.5) or via \url{http://cdsarc.u-strasbg.fr/viz-bin/cat/?J/A+A/}}}
\titlerunning{MINCE I}

\author{
G.~Cescutti \inst{1,2,3} \and
P.~Bonifacio \inst{4} \and
E.~Caffau    \inst{4} \and
L. Monaco \inst{5}\and
M.~Franchini \inst{2} \and
L.~Lombardo \inst{4}\and
A.M.~Matas Pinto \inst{4}
F.~Lucertini \inst{5,6}\and
P.~Fran\c{c}ois \inst{4,7} \and
E.~Spitoni \inst{8,9}\and
R.~Lallement \inst{4}\and
L.~Sbordone \inst{6} \and
A.~Mucciarelli \inst{10,11} \and
M.~Spite \inst{4} \and
C.J.~Hansen \inst{12} \and
P.~Di Marcantonio \inst{2} \and
A. Ku\v{c}inskas\inst{13} \and
V. Dobrovolskas\inst{13} \and
A.J. Korn \inst{14} \and
M.~Valentini \inst{15} \and
L.~Magrini \inst{16} \and
S.~Cristallo \inst{17,18} \and
F.~Matteucci \inst{1,2,3}
}
\authorrunning{Cescutti et al.}
\institute{
Dipartimento di Fisica, Sezione di Astronomia, Università di Trieste, Via G. B. Tiepolo 11, 34143 Trieste, Italy
\and
INAF, Osservatorio Astronomico di Trieste, Via Tiepolo 11, I-34143 Trieste, Italy
\and
INFN, Sezione di Trieste, Via A. Valerio 2, I-34127 Trieste, Italy
\and
GEPI, Observatoire de Paris, Universit\'{e} PSL, CNRS,  5 Place Jules Janssen, 92190 Meudon, France
\and Departamento de Ciencias Fisicas, Faculdad de Ciencias Exactas, Universidad Andres Bello, Av. Fernandez Concha 700, Las Condes, Santiago, Chile
\and ESO - European Southern Observatory, Alonso de Cordova 3107, Vitacura, Santiago, Chile
\and UPJV, Universit\'e de Picardie Jules Verne, P\^ole Scientifique, 33 rue St Leu, 80039, Amiens, France
\and Universit\'e C\^ote d'Azur, Observatoire de la C\^ote d'Azur, CNRS, Laboratoire Lagrange, Bd de l'Observatoire,  CS 34229, 06304 Nice cedex 4, France   \and Stellar Astrophysics Centre, Department of Physics and Astronomy, Aarhus University, Ny Munkegade 120, DK-8000 Aarhus C, Denmark 
\and
Dipartimento di Fisica e Astronomia, Universit\`a degli Studi di Bologna, Via Gobetti 93/2, I-40129 Bologna, Italy
\and
INAF - Osservatorio di Astrofisica e Scienza dello Spazio di Bologna, Via Gobetti 93/3, I-40129 Bologna, Italy
\and Goethe University Frankfurt, Institute for Applied Physics, Max-von-Laue-Str. 12, 60438, Frankfurt am Main, Germany; Institute for Nuclear Physics, Technical University Darmstadt, Schlossgartenstr. 2 (S2|11), 64289, Darmstadt, Germany
\and 
Institute of Theoretical Physics and Astronomy, Vilnius University, Saul\.{e}tekio al. 3, Vilnius, LT-10257, Lithuania
\and
Observational Astrophysics, Department of Physics and Astronomy, Uppsala University, Box 516, SE-751 20 Uppsala, Sweden
\and
Leibniz-Institut für Astrophysik Potsdam (AIP), An der Sternwarte 16, 14482, Potsdam, Germany
\and INAF, Osservatorio Astrofisico di Arcetri, Largo E. Fermi 5, 50125, Firenze, Italy
\and INAF, Osservatorio Astronomico d'Abruzzo, Via Mentore Maggini snc, 64100 Teramo, Italy
\and 
INFN, Sezione di Perugia, Via A. Pascoli snc, 06123 Perugia, Italy}

\date{Received September 15, 1996; accepted March 16, 1997}

 \abstract
 {In recent years, Galactic archaeology has become a particularly vibrant field of astronomy, with its main focus set on the oldest stars of our Galaxy. 
In most  cases, these stars have been identified as the most metal-poor. However, the struggle to find these ancient fossils has produced an important bias in the observations -- in particular, the  intermediate metal-poor stars ($-$2.5$<$[Fe/H]$<-$1.5) have been frequently overlooked. The missing information has consequences for the precise study of the chemical enrichment of our Galaxy, in particular for what concerns neutron capture elements and it  will be only partially covered by future multi object spectroscopic surveys such as WEAVE and 4MOST.
}
{Measuring at Intermediate Metallicity Neutron Capture Elements (MINCE) is gathering the first high-quality spectra (high signal-to-noise ratio, S/N, and high resolution) for several hundreds of bright and metal-poor stars, mainly located in our Galactic halo.} 
{We compiled our selection mainly on the basis of {\em Gaia} data and determined the stellar atmospheres of our sample and the chemical abundances of each star.}
{In this paper, we present the first sample of 59 spectra of 46 stars. We measured the radial velocities and computed the Galactic orbits for all stars.
We found that 8 stars belong to the thin disc, 15 to disrupted satellites, and the remaining cannot be associated to  the mentioned  structures, and we call them halo stars. 
For 33 of these stars, we provide abundances for the elements up to zinc.
We also show the chemical evolution results for eleven chemical elements, based on recent models.}
{Our observational strategy of using multiple telescopes and spectrographs 
to acquire high S/N and high-resolution spectra for intermediate-metallicity  stars has proven
to be very efficient, since the present sample was acquired over only about one year
of observations. Finally, our target selection strategy, after an initial 
adjustment, proved satisfactory for our purposes.}
\keywords{Galaxy: evolution - Galaxy: formation -  Galaxy: halo  – stars: abundances - stars: atmospheres - nuclear reactions, nucleosynthesis, abundances}
\maketitle
\section{Introduction}
   The project titled Measuring at Intermediate metallicity
  Neutron-Capture Elements (MINCE) is aimed at gathering abundances
  for neutron-capture elements for several hundreds stars
  at  intermediate metallicity using different facilities worldwide.
  The main idea is to study the nucleosynthetic signatures that can be found in old stars, in particular, among 
the
  specific class of chemical 
elements with Z $>$  30, that is, the  neutron-capture elements.
  They are mainly formed through multiple
  neutron captures and not through the fusion reaction that create the vast majority of elements up to the iron peak.
 The neutron-capture process is split in the rapid process
  (r-process) or slow process (s-process) depending on whether the
  timescale for neutron capture is faster or slower than radioactive
  beta decay, according to the initial definition by \citet{BBFH57}. These elements have complex nucleosynthesis and they are
  not yet deeply investigated as, such as  $\alpha -$elements.
  Recent investigations expanded the number
  of stars with detailed chemistry at extremely low metallicity up to approximately a  thousand objects \citep[e.g.][]{roederer14,Yong14}.
  After this incredible effort in searching and measuring the most
  extreme metal-poor stars (which is still ongoing), it is natural to
  think that adding valuable knowledge in this field can be difficult
  or extremely expensive, especially in terms of observing time.  However, the search for the lowest possible
  metallicity  almost completely ignored all the stars in the
  intermediate range of metallicity between the very metal-poor stars
  ([Fe/H]$<-$2.5) and thin or thick disc stars ([Fe/H]$>-$1.5). In this
  region, the number of stars with any measurements of the neutron-capture 
  elements is small, only 25\% (332 objects) according to the
  sample gathered by the JINA database (1213) and less than 10\% (103) with Eu
  measurements. According to the metallicity distribution function of
  the Galactic halo \citep{toposVI} there are more  halo stars in this region, by a factor of 12,  than   at lower
  metallicity; therefore, an enormous number of halo stars are yet unexplored   as far as the abundances of neutron-capture elements are concerned.  
  
  That apart, the more general target of a complete 
  census of the Galactic halo stars, several scientific questions can
  be addressed thanks to new abundance measurements of neutron capture
  elements. It will be possible to study how
  the spread in the n-capture elements shrinks. The spread is
  produced by stochastic process driven by the rarity of the r-process
  events \citep{Argast04,Cescutti08}, but the way this
  dispersion shrinks at higher metallicity constrains the rate of the
  r-process events in the Galactic halo \citep{Cavallo21}. Hidden in this
  region, we could find also signatures of different types of r-process(es)
  that can have polluted the interstellar medium at different timescales. This could be the case if both neutron star mergers and
  magneto rotational driven SNe have contributed to the present amount
  of r-process material \citep{Cescutti15,Cote18,Simonetti19}.  Moreover, considering the possibility that
  a large fraction of the Galactic halo originally evolved in a
  massive satellite \citep{Haywood18,Vincenzo19,Cescutti20}, we also expect that the production of s-process
  elements by AGB stars has left a signature in the chemical
  abundances at this intermediate metallicity.
   In this initial paper, we present a first sample of 59 stellar spectra (46 unique stars). We present the
atmospheric parameters measured for 41 of them, while 5 stars show an initial estimate of \Teff $<4000$K and this temperature is outside the parameter space where we believe our stellar atmosphere models are fully reliable, so we prefer to exclude them. We perform the detailed abundances determinations only of 33 stars, with 8 stars being too metal-rich for MINCE goals. The spectra
of these stars were taken at four different facilities thanks to four accepted proposals
and it clearly shows the joint efforts of the MINCE team.
Two stars, BD+07 4625 and BD+25 4520, have spectra taken from two different facilities; we decided to carry out the determination of the stellar atmospheres and chemical abundances two times independently, to check the consistency of our method.

We also introduce how we have selected our MINCE stars
and the issues that we have found in the search of an optimal 
selection of  bright halo stars for our telescopes. 
Finally, we describe the approach we intend to assume for all the MINCE stars to determine the atmospheric parameters of the stars. 
For this first sample, the results  of the chemical abundances cover
the elements up to zinc. The actual measurement of the heavy neutron capture elements will be tackled in the next MINCE paper.
We also investigated the kinematics of the stars in our sample making use of the Gaia astrometric parameters and the radial velocities (RV) we measured.
All the results obtained and published by MINCE project will produce 
a catalogue of high-quality spectra with precise atmospheric parameters and  chemical abundances constructed by combining observations from
  several facilities. 


\section{Survey description}
The concept for the survey was initiated in February 2019, thanks to a discussion between two of the authors (GC and PB) of this manuscript. The idea was (and still is) to fill an existing statistical gap in the stellar abundance  with regard to neutron capture elements (but not exclusively) in the region between $-2.5<$[Fe/H]$<-1.5$.

The organisation of the survey is  not standard: we decided to avoid intensive applications for hundreds of stars within a single facility (or up to a few). It is a diffuse plan that allows us to use several different facilities, thanks to the large collaboration. At present, we have obtained data from more that ten facilities and possibly more will be included. We try to exploit at best the time of national infrastructures too, infrastructures at the top level in terms of resolution and quality of the spectrographs, but not with the widest collecting areas (although we did apply also for ESO-VLT time). For these reasons, our targets were selected to be bright (most of them have G$<$11), with the aim attaining a high signal-to-noise ratio (S/N). The principal investigator (PI) of the single proposal within MINCE is not always the same person, but they typically vary from one facility to the other (and from one semester to the other). 
 We also decided to select  K giant stars because they are cooler than turn-off stars and
 the lines are stronger \citep[see e.g.][]{Cayrel01}. We could have also used K dwarfs, that have the same effective temperatures as K giants, however, there are two further advantages of using K giants over K dwarfs: {\em i)} the lines of ionised species, that is, the vast majority of the lines of n-capture elements, are stronger in giants than dwarfs; {\em ii)} the K dwarfs are intrinsically faint, thus the survey volume is much smaller than when using giants -- this would make it much more difficult to find bright metal-poor K dwarfs than it is to find bright metal-poor K giants.
 
\begin{table*}
\caption{Awarded time in September 2022 by the MINCE project.}\label{tablemince}
\begin{tabular}{|l|l|l|l|l|}
\hline
  \multicolumn{1}{|c|}{telescope} &
  \multicolumn{1}{c|}{instrument} &
  \multicolumn{1}{c|}{time} &
  \multicolumn{1}{c|}{targets} &
  \multicolumn{1}{c|}{status} \\
\hline
A40-41  TNG & HARPS-N & 21 h & 31 & observed\\
A42  TNG & HARPS-N & 1n  & 12  & observed\\
A43  TNG & HARPS-N & 1n  & 16  & observed \\
    \hline
 CFHT 2019B+20A & ESPaDOnS & 30h & 12 & observed\\
 CFHT 2020B & ESPaDOnS & 24.5h & 6 & observed \\
    \hline
  OHP 2019B+20A & Sophie &   6n &  42 & observed \\
    \hline
   TBL 2020A   & NeoNArval & 13h & 12 & observed (reduction problematic) \\
\hline
2019B  2.2m & FEROS & 4n & 65(72) & observed (2n cancelled) \\
2020B  2.2m & FEROS & 2n & 65 & observed\\
    \hline
    Magellan& MIKE&  2n & 14 (20) & observed (1 night cancelled) \\
\hline
  VLT ESO period 105-107 & UVES & 50h & 50 & observed \\
  VLT ESO period 106 & UVES & 50h & 50 & observed \\
\hline
  period 61, NOT & FIES &  3n  &  16 & observed\\
  period 62, NOT & FIES &  8h  &  8  & observed   \\
\hline 
     ChETEC-INFRA 1, NOT & FIES & 3n & 0 (16) &  not taken due to eruption\\
     ChETEC-INFRA 3, NOT & FIES & 3n & 5 (16) &  bad seeing, success rate 30\%\\
      ChETEC-INFRA 5, NOT & FIES & 3n & 16 &  to be taken in Oct-Dec 2022\\

\hline
         Moletai 1.65m & VUES & 38n & 24 & observed\\
\hline
  \end{tabular}
\tablefoot{ The column "targets" list the number of target observed; between brackets the requested number, when the observation was not fully successful.}
\end{table*}

The original concept was to obtain around 1000 stars in five years. We  obtained around 400 stellar spectra (see Table \ref{tablemince}) in the first two years of submissions, which perfectly matched our timetable. However, we have decided to slow down our proposal submissions to dedicate more time for the  analysis of our data and the delivery of our results.
We note that we plan to start submitting a subsequent proposal six months from now
and we will most likely postpone the end of the survey.

Surely, present surveys such as WEAVE  \citep{weave} and 4MOST \citep{4most} will produce spectra for these stars (although some of the MINCE stars may be too bright) in this range of metallicity. Still, the
  wavelength range of the high resolution surveys for these instruments is limited and  will not deliver all the elements
  that we intend to provide as part of the MINCE project. We feel that MINCE can be seen as complementary to this huge surveys, while certainly considering the completely different means involved.

\section{Target selection\label{sel}} 

The stellar candidates were selected  to be metal-poor ([M/H]$<-$0.7) and bright (V$<$10) giants ($T_{\rm eff}<$5000 K) 
based on Starhorse \citep{Anders19}. We named this method 'mince1'. Starhorse combines the precise parallaxes and optical photometry delivered by {\em Gaia}’s second data release with the photometric catalogues of Pan-STARRS1, 2MASS, and AllWISE and derived Bayesian stellar parameters, distances, and extinctions for 137 million stars.
After the first night at the Telescopio Nazionale Galileo (TNG, details of the facility are given in Sect. \ref{TNG}) covering eight candidates, we found that
 the selection provided 
 cool giant stars (see Fig. \ref{fig:tgall}), but not the requested metallicity range: the candidates were too metal-rich ($-$0.6$<$[Fe/H]$<$0) for the MINCE goals.   
For this reason, we decide to add a constrain on the kinematics of the stars (v$_{\rm tot}>200$\,kms$^{-1}$) to select halo stars, exploiting the precise measurements of {\em Gaia}. This selection scheme improved  the success rate to 100\%: all the stars present [Fe/H]$<-$1.4. We named this method 'mince2'. The  eight stars mentioned above are not fully considered here,  given their metallicities are above the threshold we set for MINCE, and we present only their atmospheric parameters; the analysis of their chemical abundances will be carried out in a forthcoming paper devoted to more metal rich stars compared to MINCE limits. 
 The sample comprises relative bright objects and we set the observations to   approximately reach S/N $\sim$ 100  at 500\,nm.
We also include two stars that were actually selected from the Apache Point Observatory Galactic Evolution
Experiment (APOGEE) survey \citep{apogee}. With a higher resolution and different spectral coverage, MINCE can provide different elements and also a comparison with the results obtained by APOGEE in the infrared.

\section{Observations and data reduction} 

\begin{figure}[h]
\centering
\includegraphics[width=\hsize,clip=true]{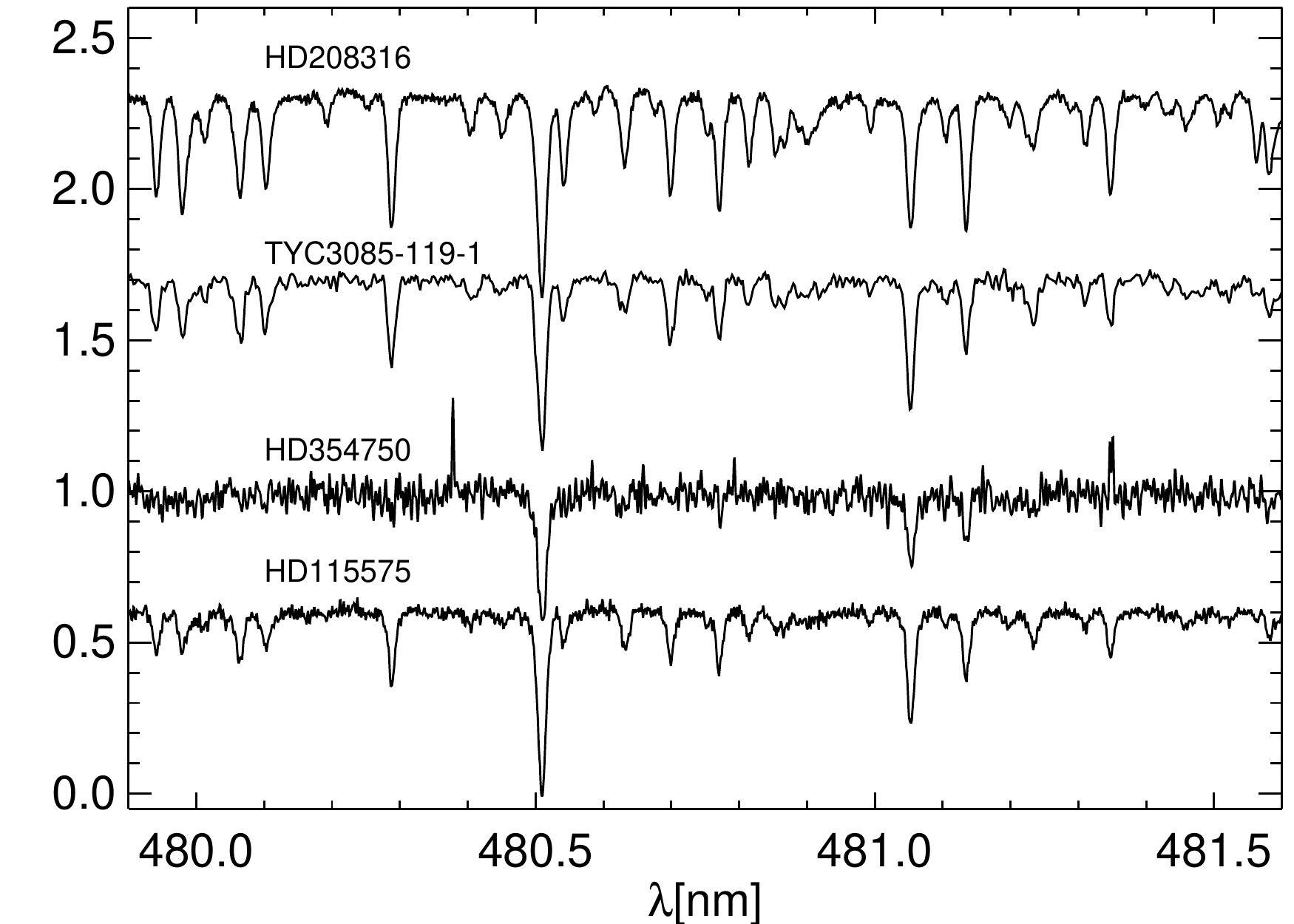}
\caption{Example of the spectra acquired to give an idea of the quality.
The spectral region around the \ion{Zn}{i} 481.0\,nm line. The normalised spectra
have been shifted vertically for display purposes.}
\label{fig:spectra}
\end{figure}

As mentioned in the introduction, this sample comprises spectra taken from several facilities and obtained thanks to a total of four proposals with three 
different PIs: Cescutti for HARPS-north at TNG, E. Spitoni for FIES at NOT
and P. Bonifacio for Sophie at OHP, and ESPaDOnS at CFHT.
Details on the observations are provided in Tables \ref{harpsn} to \ref{espadons}. An example of the spectra acquired is shown in Fig.\,\ref{fig:spectra}, where two spectra have the desired S/N ($\sim100$ at 550nm, the first two from the top); clearly, during an observational campaign not everything is perfect and indeed the other two spectra present a S/N lower, in particular,  HD 354750 with a S/N$\sim$50. 

\subsection{TNG HARPS-N}\label{TNG}
 The 3.58m telescope Telescopio Nazionale Galileo (TNG), is the Italian facility located at the Roque de Los Muchachos Observatory in the Canary island of La Palma. 
 We used HARPS-N \citep{HARPSN}, which is a high-resolution (resolving power R=115,000), high-stability visible (383-693 nm) spectrograph.  Long-term stability allows an accuracy better than $<$1ms$^{-1}$ in the radial velocity measurements and it is excellent for the discovery and characterisation of exoplanets, but it is also well suited for stellar abundance spectroscopy.
The spectra were  taken in service mode  in two nights in May and June 2020. 
For the reduction of the echelle spectra, we used the standard pipeline.
The radial velocities are determined by the pipeline through cross-correlation with a mask that is appropriate for the spectral type of the star.

\subsection{NOT FIES}
\begin{figure}[h]
\centering
\includegraphics[width=\hsize,clip=true]{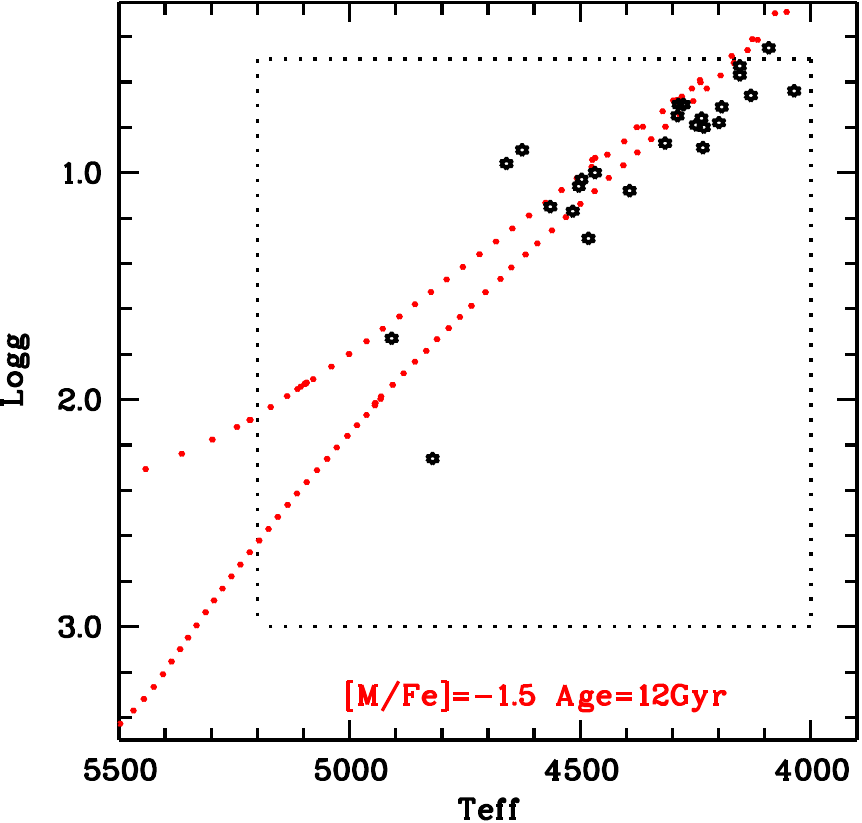}
\caption{ \teff,\logg\  plot with the observed stars here analysed (black open stars) and 
a PARSEC \citep{Leo,marigo17} isochrone
of 10\,Gyr and metallicity $-$1.5 (red dots). 
The dotted lines delimit the grid used by \mygi\ in the chemical analysis.}
\label{fig:tgall}
\end{figure}

The Nordic Optical Telescope (NOT) is a 2.56-m telescope also located at the Spanish Roque de los Muchachos Observatory, about 1km away from the TNG. 
We used  FIES \citep{FIES},  a cross-dispersed high-resolution echelle spectrograph with a maximum spectral resolution of R = 67000. The entire spectral range 370-830 nm is covered without gaps in a single, fixed setting.  Most of the spectra were taken in service mode during  June 2020 (see Table \ref{fies} for specific dates). 
Also in this case, we used the output of the standard pipeline, which are available upon request.
Radial velocities are not provided by the pipeline. They have been determined by
template matching \citep[see e.g.][]{2011ApJ...736..146K} over the range between 400\,nm to 660\,nm. The template was a synthetic spectrum computed with the parameters 
provided in Table \ref{fig:tgall}.
The error provided in Table\,\ref{fies} is just based on the $\chi^2$ of the fit and does not take into account systematic errors. The
systematic errors due to the fact that the calibration arc was taken
several hours before the observation can be of the order of a few 100 m$^{-1}$
(J. Telting, priv. comm.).
The mid-exposure time was taken
from the descriptor DATE-AVG in the FITS header of each observation. From this time,
the Barycentric Julian Date (BJD) and the  
barycentric correction were computed  using the tools OSU Online Astronomy Utilities\footnote{\url{https://astroutils.astronomy.osu.edu/}} that implement the methods and algorithms described in \citet{2014PASP..126..838W}.

\subsection{OHP 1.93 Sophie}

The OHP 1.93m telescope is located in at the Observatoire de Haute Provence, in  southern France. 
The spectra were obtained with the Sophie spectrograph \citep{2006tafp.conf..319B} in high-resolution mode, providing a resolving power R=75 000 and a spectral range
from 387.2 nm to 694.3 nm. The spectra were obtained in visitor
mode, over three nights from August 24 to 26, 2020, and the
observer was P. Bonifacio. 
The wavelength calibration relied both on a Th-Ar lamp 
and on a Fabry-Pérot etalon. The data were reduced automatically 
on-the-fly by the Sophie pipeline. In a similar way as for HARPS-N, the pipeline determines radial velocities
from cross-correlation with a suitable mask.

\subsection{CFHT ESPaDOnS}

The 3.6m Canada-France-Hawaii telescope (CFHT) is located on
the summit of Mauna Kea, on the island of Hawai, USA. The spectra were obtained with the   fibre-fed spectropolarimeter ESPaDOnS \citep{Espadons}. The observation were obtained in the queued service observation mode of CFHT in 2020. The spectroscopic mode “Star+Sky” was used, providing a resolving power of R=65 000 and the spectral range 370 nm to 1051 nm. The data was delivered to us reduced with
the Upena\footnote{\url{http://www.cfht.hawaii.edu/Instruments/Upena/}} 
pipeline that uses the routines of the Libre-ESpRIT
software \citep{1997MNRAS.291..658D}. 
The output spectrum is provided in
an order-by-order format, we merged the orders using an ESO-MIDAS\footnote{\url{https://www.eso.org/sci/software/esomidas/}}
with a script written by ourselves.
The pipeline applies the barycentric correction to the reduced spectrum and provides the Heliocentric Julian Date (HJD),
we transformed this to BJD using a specific tool 
\footnote{\url{https://astroutils.astronomy.osu.edu/time/utc2bjd.html}} that implements the methods and algorithms described in 
\citet{2014PASP..126..838W}.
The pipeline corrects the wavelength scale using
the telluric absorption lines and this should compensate
for the difference in temperature and pressure
between the time when the calibration arc was taken
and the time of the observation.
As for the FIES spectra we measured the radial velocities by
template matching over the range 400\,nm to 660\,nm. 
We underline that the error provided in Table \ref{espadons}
is based on the $\chi^2$ of the fit, thus taking into account only the noise in the spectrum and not any systematic error. 
In spite of the fact that ESPaDOnS is protected by
two thermal enclosures, its temperature and pressure are
not actively controlled, just as those of HARPS-N or Sophie. 
According to the 
documentation 
the expected precision using the telluric correction is 20 ms$^{-1}$. 
For star 
TYC 3085-119-1 we have two spectra, but although the radial velocity
was measured for both, the chemical analysis was performed only on the
second spectrum that has S/N $\sim$ 100 at 500\,nm against $\sim$ 60 for the first spectrum. The improvement in S/N by coadding the two spectra would be marginal.

\subsection{Radial velocities}

Our measured radial velocities are generally in very good agreement
with the {\em Gaia} DR3 radial velocities \citep{KatzDR3}. 
However there are twenty spectra, with sixteen stars
for which the difference between our measurement and {\em Gaia} DR3
exceeds $3 \sigma$, where $\sigma$ is computed by adding under quadrature
the errors associated to each measurement.
In some cases, this is certainly due to real radial velocity variations
and this may be because the star is in a binary system.
In none of our spectra we detect a secondary spectrum of a companion, 
so if any of the stars is binary indeed, the companion must be much less
luminous, implying a small veiling of the spectrum. This gives us confidence on our approach of analysing all stars as single stars.

The most clear case is TYC 4584-784-1 for which our Sophie radial velocity differ by 7.18 \kms\ from that of {\em Gaia} Dr3. 
Also the error on the {\em Gaia} radial velocity is large for a star of this brightness, 2.34\,\kms .
Other very clear cases are TYC 4331-136-1, BD\,--07\,3523,
BD\,+24\,2817, HD\,139423, HD\,354750, and
TYC\,565-1564-1.
A borderline case is that of TYC\,2824-1963-1 
two Sophie spectra provide radial velocities that differ by just over 3$\sigma$ from that of {\em Gaia}, which, however, has a an error of only 0.28\kms .

A controversial case is put forth by  BD\,+07\,4625. For this star, we observed two spectra: with FIES and with Sophie. The Sophie radial velocity  is at 4$\sigma$ of the {\em Gaia} one, while the two FIES radial velocities are at 4. and 1.5 $\sigma$ from the {\em Gaia} one, which has a small error of
0.13 \kms.
The FIES spectra were taken 55.75 days before the Sophie ones. 
It is also useful to consider that the standard deviation of the FIES and Sophie
radial velocities is of 0.48\,kms and the mean is perfectly consistent
with {\em Gaia}. 
Another suspicious case is BD\,+25\,4520. This star
has been observed with HARPS-N and 77 days later with Sophie.
While the radial velocity derived from the Sophie agrees with the {\em Gaia}
radial velocity to better than $1 \sigma,$ with regard to the HARPS-N spectrum it differs by almost $2 \sigma$. It is interesting to note that the {\em Gaia} radial velocity has 
changed by about 1\,\kms\ from DR2 to DR3,
and that the 
{\em Gaia} error, 0.42\,\kms , is very similar to the standard
deviation of the two HARPS-N and SOPHIE measurements, 0.35\,\kms .
We suspect this star to be a radial velocity variable of low amplitude, possibly on the order of 1\,\kms .


\section{Analysis}

\subsection{Stellar parameters} \label{sec:param}

The sample presented here is the the first of a series and we expect to have many spectra to be analysed in the future (as explained in Sect. \ref{sel}).
We then need a way to analyze these stars that is as automated and objective as possible.

The stellar parameters were derived from the photometry and the parallax, by using the {\em Gaia} data release early three ({\em Gaia}\,EDR3).
We first dereddened the {\em Gaia} photometry by using the maps from \citet{schlafy}
and by iterating the computation of the dereddening we took into account
the stellar parameters.
With the dereddened G magnitude, we derived the absolute $G$ magnitude by applying the parallax, corrected for the zero point as suggested by \citet{lindegren20}.
The {\em Gaia} $G_{BP} - G_{RP}$ dereddened colour, the absolute $G$ magnitude with a first guess metallicity have been compared to synthetic colours in order to derive the first guess stellar parameters.

This first guess of the stellar parameters are  fed to \mygi\ \citep{Sbordone14} to derive the metallicity.
The metallicity derived by \mygi\ was then used as input to derive new stellar parameters, from the photometry and parallax, as described above.
\mygi repeats this process until the changes in stellar parameters were less than 10\,K in \Teff, and less than 0.05\,dex in \logg.
For the micro-turbulence, we used the calibration by \citet{mashonkina17} at any iteration and applied these values as the final choice.
The stellar parameters and derived metallicity are reported in Table\,\ref{T1}.

\begin{table*}
\caption{Stellar parameters of the sample with extra information.}

\label{T1}
\begin{tabular}{lrrrrlrrrrrr}
\hline
  \multicolumn{1}{c}{Star} &
  \multicolumn{1}{c}{\teff} &
  \multicolumn{1}{c}{\logg} &
  \multicolumn{1}{c}{$\xi$} &
  \multicolumn{1}{c}{[Fe/H]} &
  \multicolumn{1}{c}{Selection} &
  \multicolumn{1}{c}{parallax} &
  \multicolumn{1}{c}{$G$} &
  \multicolumn{1}{c}{$G_{BP}-G_{RP}$} &
  \multicolumn{1}{c}{teff50} &
  \multicolumn{1}{c}{logg50} &
  \multicolumn{1}{c}{vel$_{tot}$} \\
  \multicolumn{1}{c}{} &
  \multicolumn{1}{c}{[K]} &
  \multicolumn{1}{c}{[gcs]} &
  \multicolumn{1}{c}{\kms} &
  \multicolumn{1}{c}{} &
  \multicolumn{1}{c}{} &
  \multicolumn{1}{c}{mas} &
  \multicolumn{1}{c}{mag} &
  \multicolumn{1}{c}{mag} &
  \multicolumn{1}{c}{[K]} &
  \multicolumn{1}{c}{[cgs]} &
  \multicolumn{1}{c}{\kms} \\
\hline
  TYC    4-369-1 & 4234 & 0.89 & 1.94 & -1.84 & mince2 & 0.216 & 10.78 & 1.49 & 4439 & 0.96 & 345.0\\     
  BD+04    18 & 4053 & 0.74 & 1.9 & -1.48 & mince2 & 0.293 & 9.19 & 1.6 & 4284 & 0.67 & 484.0\\           
  TYC   33-446-1 & 4289 & 0.75 & 2.07 & -2.22 & mince2 & 0.185 & 10.09 & 1.52 & 4323 & 0.69 & 280.0\\     
  TYC 2824-1963-1 & 4036 & 0.64 & 1.95 & -1.6 & mince2 & 0.185 & 10.06 & 1.69 & 4241 & 0.66 & 433.0\\     
  TYC 4331-136-1 & 4133 & 0.5 & 2.13 & -2.53 & mince2 & 0.513 & 9.53 & 2.11 & 4385 & 0.74 & 201.0\\       
  HD  87740 & 4746 & 1.89 & 1.62 & -0.56 & mince1 & 1.448 & 8.56 & 1.2 & 4838 & 1.96 & 45.0\\             
  BD+31  2143 & 4565 & 1.15 & 2.03 & -2.37 & mince2 & 0.595 & 8.87 & 1.3 & 4689 & 1.26 & 359.0\\          
  HD  91276 & 4610 & 1.73 & 1.63 & -0.58 & mince1 & 1.372 & 8.57 & 1.26 & 4802 & 1.83 & 60.0\\            
  BD+13  2383 & 4458 & 1.54 & 1.65 & -0.56 & mince1 & 1.319 & 8.54 & 1.35 & 4751 & 1.68 & 27.0\\          
  BD-07  3523 & 4193 & 0.71 & 2.02 & -1.95 & mince2 & 0.408 & 9.12 & 1.58 & 4410 & 0.83 & 249.0\\         
  HD 115575 & 4393 & 1.08 & 1.94 & -1.99 & mince2 & 0.694 & 9.02 & 1.45 & 4579 & 1.26 & 324.0\\           
  BD+48  2167 & 4468 & 1.0 & 2.04 & -2.29 & mince2 & 0.429 & 9.32 & 1.36 & 4498 & 1.09 & 255.0\\          
  BD+06  2880 & 4167 & 0.82 & 1.91 & -1.45 & mince2 & 0.616 & 9.18 & 1.53 & 4463 & 1.14 & 378.0\\         
  BD+32  2483 & 4516 & 1.17 & 1.99 & -2.25 & mince2 & 0.404 & 9.83 & 1.32 & 4473 & 1.3 & 259.0\\          
  HD 130971 & 4045 & 1.21 & 1.61 & -0.64 & mince1 & 1.247 & 8.6 & 1.68 & 4658 & 1.48 & 40.0\\             
  BD+24  2817 & 4722 & 1.89 & 1.56 & 0.02 & mince1 & 1.61 & 8.54 & 1.3 & 4981 & 2.12 & 73.0\\             
  HD 238439 & 4154 & 0.53 & 2.1 & -2.09 & mince2 & 0.29 & 9.26 & 1.6 & 4533 & 0.96 & 415.0\\              
  HD 138934 & 4725 & 2.41 & 1.34 & -0.19 & mince1 & 2.296 & 8.01 & 1.26 & 4947 & 2.12 & 27.0\\            
  HD 139423 & 4287 & 0.7 & 2.05 & -1.71 & mince2 & 0.808 & 8.02 & 1.5 & 4369 & 0.92 & 431.0\\             
  HD 142614 & 4316 & 0.87 & 1.96 & -1.46 & mince2 & 0.668 & 8.73 & 1.45 & 4370 & 1.12 & 412.0\\           
  BD+11  2896 & 4254 & 1.07 & 1.83 & -1.41 & mince2 & 0.771 & 8.72 & 1.48 & 4243 & 1.21 & 286.0\\         
  BD+20  3298 & 4154 & 0.57 & 2.07 & -1.95 & mince2 & 0.476 & 8.77 & 1.64 & 4742 & 1.39 & 423.0\\         
  TYC 2588-1386-1 & 4130 & 0.66 & 1.99 & -1.74 & APOGEE & 0.129 & 11.73 & 1.58 & 4319$^*$ & 1.27$^*$ & 289.0\\
  TYC 3085-119-1 & 4820 & 2.26 & 1.56 & -1.51 & APOGEE & 0.954 & 10.38 & 1.12 & 4745$^*$ & 2.14$^*$ & 122.0\\
  BD+39  3309 & 4909 & 1.73 & 1.94 & -2.58 & mince2 & 0.704 & 9.6 & 1.1 & 4855 & 1.9 & 300.0\\            
  HD 165400 & 4942 & 1.68 & 1.79 & -0.25 & mince1 & 1.37 & 8.34 & 1.27 & 4825 & 1.78 & 23.0\\             
  TYC 1008-1200-1 & 4199 & 0.78 & 2.01 & -2.23 & mince2 & 0.226 & 10.19 & 1.74 & 4335 & 0.7 & 426.0\\     
  TYC 4221-640-1 & 4295 & 0.66 & 2.12 & -2.27 & mince2 & 0.188 & 10.59 & 1.55 & 4421 & 0.82 & 387.0\\     
  TYC 4584-784-1 & 4232 & 0.8 & 2.0 & -2.04 & mince2 & 0.192 & 10.62 & 1.59 & 4261 & 0.78 & 326.0\\       
  TYC 3944-698-1 & 4091 & 0.45 & 2.11 & -2.18 & mince2 & 0.225 & 9.9 & 1.81 & 4523 & 0.96 & 270.0\\       
  HD 354750 & 4626 & 0.9 & 2.17 & -2.36 & mince2 & 0.177 & 10.59 & 1.43 & 4426 & 0.94 & 235.0\\           
  BD-00  3963 & 4970 & 1.92 & 1.68 & -0.13 & mince1 & 1.68 & 8.54 & 1.3 & 4936 & 2.0 & 43.0\\             
  BD+07  4625$^a$ & 4757 & 1.64 & 1.86 & -1.93 & mince2 & 1.209 & 8.61 & 1.24 & 4877 & 1.79 & 570.0\\     
  BD+07  4625$^b$ & 4757 & 1.64 & 1.86 & -1.95 & mince2 & 1.209 & 8.61 & 1.24 & 4877 & 1.79 & 570.0\\     
  BD+25  4520$^b$ & 4276 & 0.7 & 2.08 & -2.28 & mince2 & 0.245 & 9.25 & 1.61 & 4386 & 0.72 & 445.0\\      
  BD+25  4520$^c$ & 4276 & 0.7 & 2.08 & -2.27 & mince2 & 0.245 & 9.25 & 1.61 & 4386 & 0.72 & 445.0\\      
  HD 208316 & 4249 & 0.79 & 1.98 & -1.61 & mince2 & 0.654 & 8.35 & 1.51 & 4390 & 0.9 & 315.0\\            
  TYC 4267-2023-1 & 4660 & 0.96 & 2.11 & -1.74 & mince2 & 0.62 & 9.5 & 1.84 & 4607 & 1.16 & 372.0\\       
  BD+21  4759 & 4503 & 1.06 & 2.05 & -2.51 & mince2 & 0.397 & 9.44 & 1.37 & 4565 & 1.15 & 266.0\\         
  BD+35  4847 & 4237 & 0.76 & 2.01 & -1.92 & mince2 & 0.644 & 8.46 & 1.61 & 4725 & 1.48 & 263.0\\         
  BD-00  4538 & 4482 & 1.29 & 1.88 & -1.9 & mince2 & 0.853 & 8.77 & 1.34 & 4607 & 1.41 & 320.0\\          
  TYC 4001-1161-1 & 4129 & 0.75 & 1.94 & -1.62 & mince2 & 0.42 & 10.09 & 1.87 & 4556 & 1.07 & 423.0\\     
  BD+03  4904 & 4497 & 1.03 & 2.06 & -2.58 & mince2 & 0.398 & 9.5 & 1.38 & 4528 & 1.1 & 307.0\\           
  \hline
  \\
  \multicolumn{4}{l}{$^a$ FIES spectrum}\\
  \multicolumn{4}{l}{$^b$ SOPHIE spectrum}\\
\multicolumn{4}{l}{$^c$ HARPS-N spectrum}\\
  \end{tabular}
\tablefoot{For the column labeled 'selection', see Sect. \ref{sel}; columns labeled parallax, $G$, and $G_{BP} -G_{RP}$ are from {\em Gaia} (DR2), teff50, and log50 are from the Starhorse database (excluding *, which is taken from the APOGEE survey); vel$_{tot}$ was computed from {\em Gaia} (DR2) data considering proper motions, parallax, and radial velocity.} 
\end{table*}

\subsection{Comparison of the stellar parameters with Starhorse}
\begin{figure}[h]
\centering
\includegraphics[width=\hsize,clip=true]{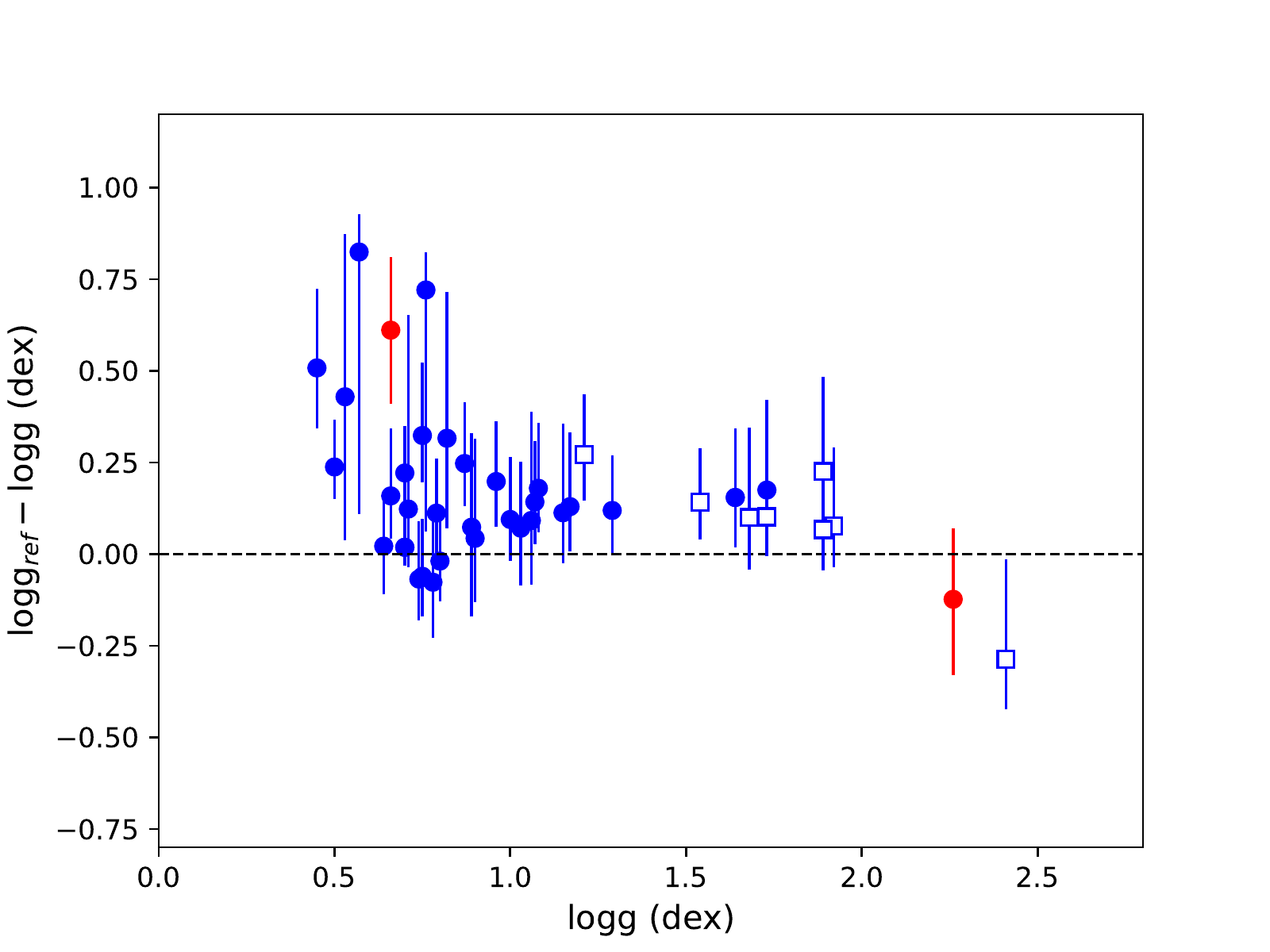}
\caption{Comparison of the \logg \, obtained for our MINCE stars to the
\logg \, obtained by the reference databases. For most of the stars, the reference is \cite{Anders19}. They are shown in blue symbols:  filled dots
for stars selected with the mince2 selection,  open squares for those selected with mince1.
The error-bars considered are  84th and 16th percentile obtained from the Bayesian approach used in Starhorse. For two stars, shown in red, the reference database for the \logg and its error is the APOGEE survey database.}
\label{fig:logg}
\end{figure}

\begin{figure}[h]
\centering
\includegraphics[width=\hsize,clip=true]{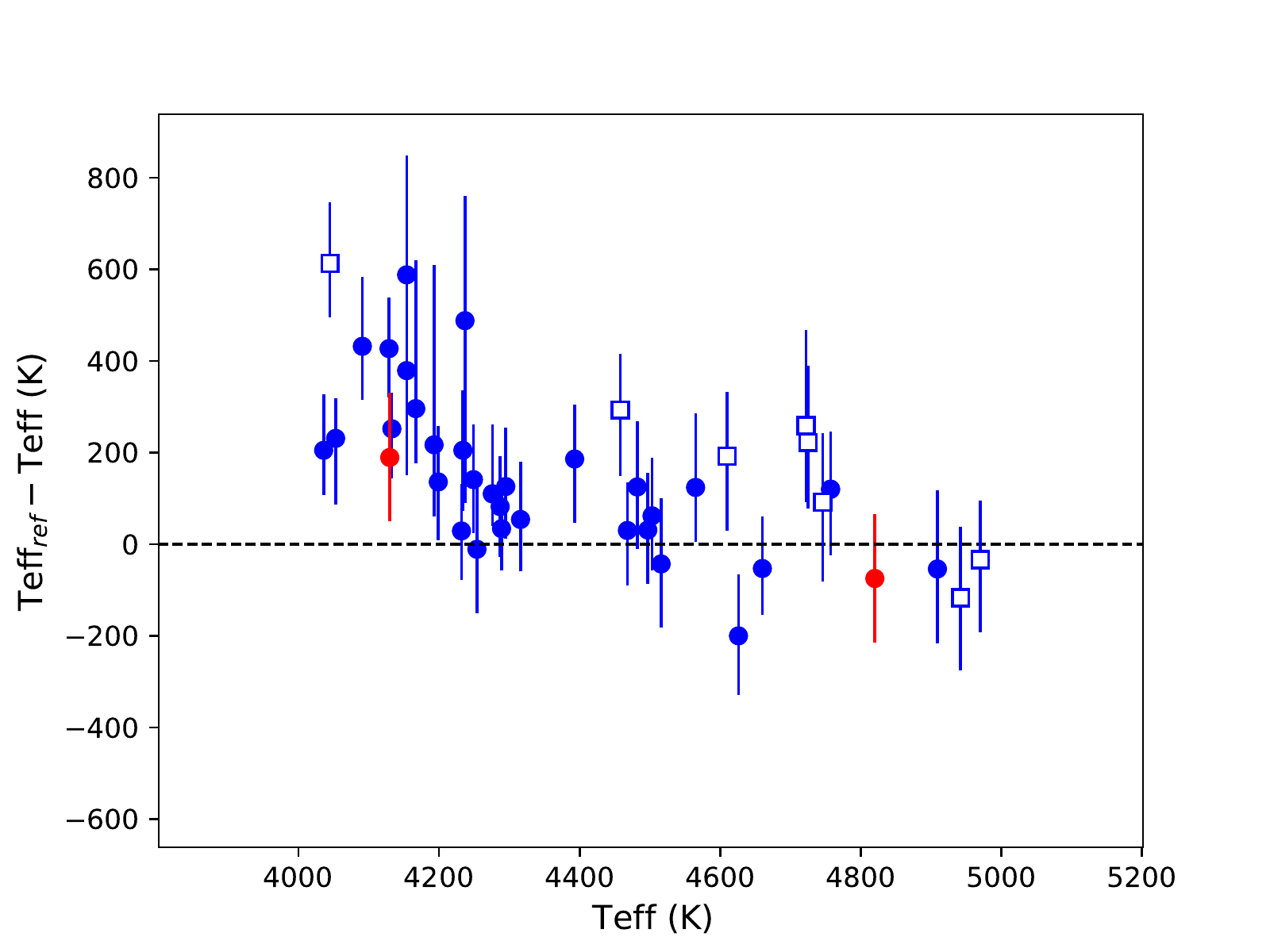}
\caption{Comparison of the \Teff \, obtained for our MINCE stars to the
\Teff \, obtained by the reference databases. 
The details are the same as in Fig.\ref{fig:logg}.}
\label{fig:Teff}
\end{figure}

The results obtained for the stellar parameters from our spectra can be compared to those obtained by Starhorse. This comparison  can be important to evaluate the use of this database. 
In Fig.\ref{fig:logg}, we show the case of \logg \, where  a positive offset (0.16 dex) and a dispersion are visible, in particular, for \logg$<$1, with a standard deviation of 0.20 dex. 
In Fig.\,\ref{fig:Teff}, we present the case of \Teff. Again, there
is a positive offset of 154 K, with a standard deviation of 176 K that is most prominent for \Teff$<$4300 K.
 Overall, the Starhorse database concerning the metallicities is not good  enough (as mentioned in Sect. \ref{sel}), but it is certainly suitable for selecting giant stars. For this reason, in the future, we will also consider to use the values derived by Starhorse as first guess of the stellar parameters \Teff \, and \logg\, applying suitable corrections inferred by the comparison with our results, omitting the procedure described above.
In both figures, we present also the comparison to the measurements obtained by the APOGEE survey, although only for a sample of two stars. We cannot draw valid conclusions from only two objects, but clearly for the cooler star the difference in \logg\, is not negligible, although the \Teff \, shows only a moderate difference of $\sim$ 200 K.

\subsection{Kinematics}\label{sec:kin}
We investigated the kinematics of the stars in our sample making use of the {\em Gaia}\,EDR3 astrometric parameters and radial velocities (RV) we measured. In particular, for the RVs we adopted weighted means for stars having SOPHIE, HARPS-N, ESPADONS, and Gaia values. In the case of ESPADONS RVs, we adopted an error on the RV of 20 ms$^{-1}$. For stars having FIES and Gaia measures or with measures more than 3 $\sigma$ different from the Gaia values, we adopted the non-weighted means of the values available. 

In order to evaluate kinematic quantities and the actions, we used the {\tt galpy} code \citep[][]{bovy15} together with the MWPotential14 potential and the solar motion of \citet[][]{schoenrich10}. We adopted a solar distance from the Galactic Centre of 8\,kpc and a circular velocity at the solar distance of 220\,\kms \citep[][]{bovy12}.

Following \citet[][]{bonifacio21}, we evaluated the corresponding errors using the {\tt pyia} code. For each stars, we extracted 1000 instances of the input parameters (coordinates, proper motions, distance, and radial velocity) from a multivariate Gaussian, which takes into account the covariance matrix. Each instance is then fed to {\tt galpy}. For each parameter calculated, we adopt as errors the standard deviations of the 1000 realisations. We notice that the quantities reported in Tables\,\ref{kinematics1} and \,\ref{kinematics2} are the value obtained from the input parameters taken at face value, only the reported errors are evaluated with the procedure described above.

\begin{figure*}[h]
\centering
\includegraphics[width=\hsize,clip=true]{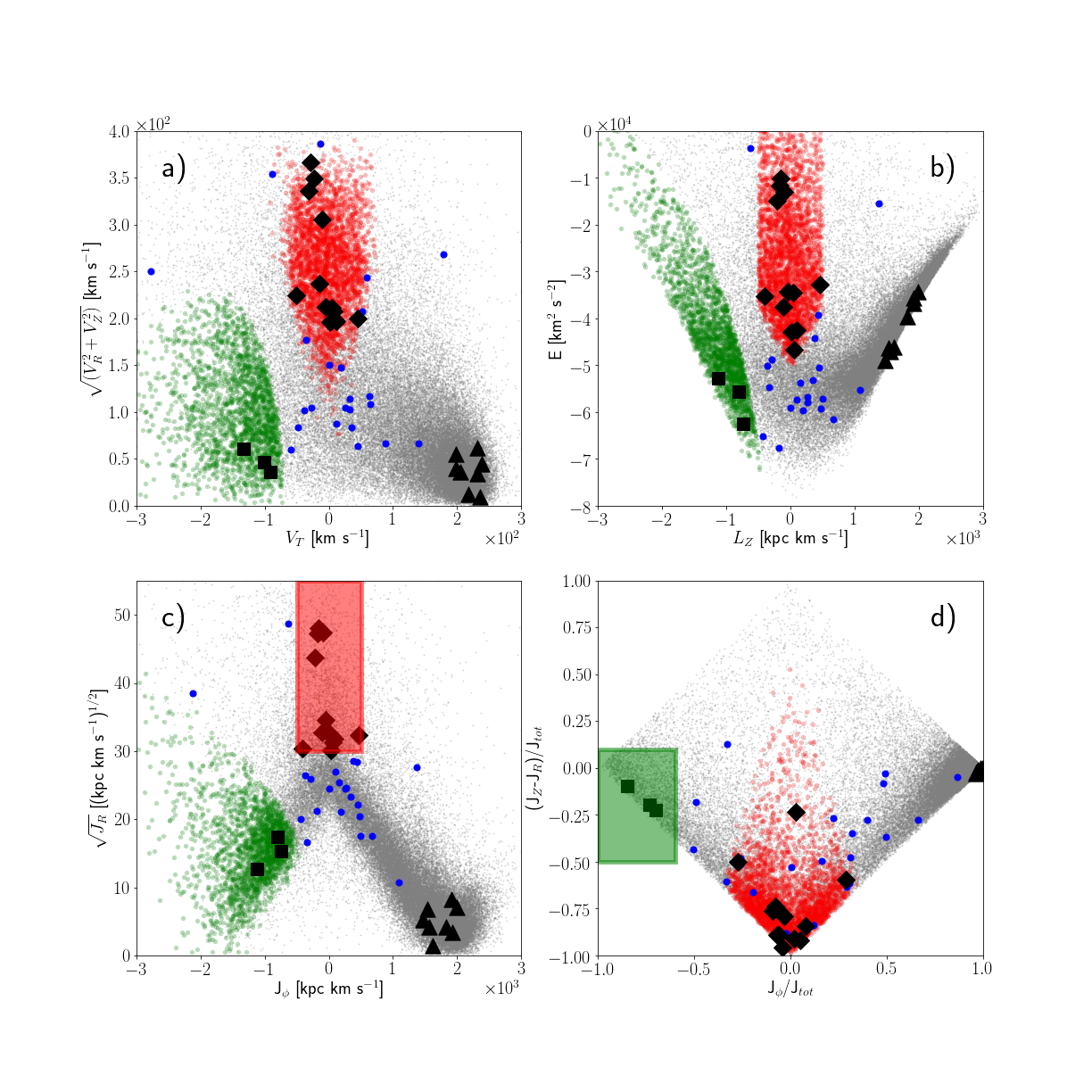}
\caption{Target stars kinematic properties are presented in various planes. Program stars are presented as big filled symbols. The background population (gray points) is the ``good parallax sample" of \citet[][]{bonifacio21} for reference. Red filled points in all planes are stars selected from the background population as likely GSE members according to the red shaded box of panel c. Green filled points are likely Sequoia members according to the green shaded box in panel d. Panel a: Toomre diagram ($\sqrt{V_R^2 + V_Z^2}$ versus V$_T$). Panel b: Orbital energy E {\it vs} angular momentum, L$_Z$. Panel d (action diamond diagram): the difference in the vertical and radial actions (J$_Z$-J$_R$) {\it vs} the azimuthal action (J$_\phi$=L$_Z$). Quantities are normalised to the total action J$_{tot}$=J$_Z$+J$_R$+|J$_\phi$|. Panel c): $\sqrt{J_R}$ {\it vs} L$_Z$. Target stars likely belonging to GSE and Sequoia are marked as black-filled diamond and squares, respectively. Targets likely belonging to the thin disc are presented as black filled triangles. The remaining targets are presented as blue filled circles.}
\label{kin}
\end{figure*}

In Figure\,\ref{kin}, we present our targets (black filled symbols) in four planes commonly used to characterise the stellar kinematic. For reference, we plot in gray in each panel the ``good parallax" sample from \citet[][]{bonifacio21}.  This is the sample of TO stars   that have parallaxes of $\varpi > 3 \times \Delta\varpi$. It is useful as high-quality reference sample to see the Galactic dynamics in action space.

Considering Galactocentric cylindrical coordinates, in panel c) we present the component of the velocity in the plane of the Galaxy (V$_T$) versus a combination of the radial and vertical component of the velocity ($\sqrt{V_R^2+V_Z^2}$), namely, a version of the Toomre diagram. In this plane, disc stars are visible as the roughly circular concentration of stars in the bottom-right of the figure. Stars in prograde motions are found at V$_T>$0.
The  panel b) shows the relation between two integrals of motion, namely the orbital energy per unit mass E versus the vertical component of the angular momentum, L$_Z$. In this plane, disc stars are found as a concentration in the middle right part of the figure. Stars in prograde motions are found at L$_Z>$0.

The panel d) shows the so-called action diamond, namely, the difference between the vertical and radial actions (J$_Z$-J$_R$) versus the azimuthal action (J$_\phi$ = L$_Z$), normalised to J$_{tot}$ = $|J_{\phi}|$+J$_Z$+J$_R$. In this plane, disc stars are found in the middle-right corner of the figure. Finally,  panel c) presents the square root of the radial action, $\sqrt{J_R,}$ {\it } versus the azimuthal action, J$_\phi$. Disc stars are visible at the bottom right of the figure. 

The panels c) and d)  were used by \citet[][]{feuillet21} to select candidates likely belonging to the {\em Gaia}-Sausage-Enceladus \citep[GSE,][]{Belokurov18,Haywood18,Helmi18} and the Sequoia \citep[][]{barba19,villanova19,myeong19} accretion events. The selection box they used for GSE and Sequoia (red and green shaded areas in panels c and d) are indicated. Stars in the background populations following in these two selections box are reported in red (GSE candidates) and green (Sequoia candidates) in panels a and b. 

GSE candidates are highly eccentric (large $\sqrt{J_R}$ values) and do not have a large angular momentum (small values of J$_\phi$ = L$_Z$). Sequoia candidates are highly retrograde (highly negative values of J$_\phi$ = L$_Z$) and their orbits are not as eccentric as those of GSE candidates. 

Among the stars in our sample, we identify 12 and 3 stars with kinematics compatible with the GSE (black-filled diamond) and Sequoia (black-filled squares) structures, respectively, according to the selections boxes of Fig.\,\ref{kin}. We identify them in Tables\,\ref{kinematics1} and \,\ref{kinematics2}. We also identified eight stars with thin disc kinematics (black filled triangles). These are the most metal-rich stars in the sample ([Fe/H]$>$-0.7\,dex) and are confined to the disc (Z$<$0.8kpc and Z$_{max<}$0.9kpc). One star likely belongs to the thick disc (TYC\,3085-119-1, filled blue circle at V$_T$=+140\,\kms in the top-left panel; [Fe/H]=-1.5, e=0.36, Z=0.6\,kpc, Z$_{max}$=1.6\,kpc). The remaining stars (filled blue circles) may be associated with the halo and have in roughly equal number prograde and retrograde orbits. They are formally not associated with GSE or Sequoia, although some of them lie quite close to GSE or Sequoia stars in the various planes. 

Shall we use the classification scheme of \citet[][]{bensby14}, besides TYC\,3085-119-1, 
we would also classify HD 143348 and HD 354750 as thick-disc stars, while the classification of the remaining stars as belonging to the thin disc or the halo would be confirmed. As expected, candidate GSE and Sequoia stars would be classified as halo stars. HD143348 and HD354750 are indeed confined to the disc (Z$_{max}$=1.19 and 2.51 kpc, respectively). They have, however, highly eccentric orbits (ecc.=0.62 and 0.87, respectively). 

\begin{table*}
 \caption{Kinematic properties of the sample.}
  
\label{kinematics1}
\hspace{-0.6cm}
\begin{tabular}{lrrrrrrr}
\hline
Star & V$_R$ & V$_T$ & V$_Z$ & r$_{ap}$ & r$_{peri}$ & ecc. & Z$_{max}$\\
& km/s & km/s & km/s & kpc & kpc & & kpc\\

\hline
TYC\,4-369-1$^b$     &  194.02 $\pm$   9.62 & -5.07 $\pm$ 18.12 & -86.58 $\pm$  5.91 &  0.10 $\pm$  0.26  & 16.10 $\pm$  1.37 &  0.99 $\pm$  0.03 &  5.88 $\pm$  0.75\\
BD\,+04\,18$^b$      &  190.95 $\pm$   7.09 & 11.10 $\pm$  9.07 & -48.12 $\pm$  3.20 &  0.20 $\pm$  0.14  & 13.29 $\pm$  0.53 &  0.97 $\pm$  0.02 &  2.58 $\pm$  0.20\\
TYC\,33-446-1        &  -16.17 $\pm$   1.51 & 45.68 $\pm$  8.91 & 61.42 $\pm$  1.57 &  1.42 $\pm$  0.28  & 10.35 $\pm$  0.13 &  0.76 $\pm$  0.04 &  4.93 $\pm$  0.20\\
TYC\,2824-1963-1$^b$ &  185.87 $\pm$   5.74 & 45.06 $\pm$ 11.94 & -73.40 $\pm$  3.13 &  1.08 $\pm$  0.29  & 16.57 $\pm$  0.65 &  0.88 $\pm$  0.04 &  4.87 $\pm$  0.60\\
TYC\,4331-136-1      &  -4.74 $\pm$   2.54 & 35.36 $\pm$  4.54 & 83.62 $\pm$  3.38 &  0.89 $\pm$  0.11  &  9.81 $\pm$  0.08 &  0.83 $\pm$  0.02 &  4.53 $\pm$  0.42\\
HD\,87740$^c$        &  -62.00 $\pm$   0.56 & 232.26 $\pm$  0.21 & -2.20 $\pm$  0.12 &  7.20 $\pm$  0.01  & 11.36 $\pm$  0.02 &  0.22 $\pm$  0.00 &  0.61 $\pm$  0.01\\
BD\,+31\,2143$^a$    &  -22.14 $\pm$   0.69 & -91.74 $\pm$  6.53 & 28.40 $\pm$  0.71 &  2.33 $\pm$  0.22  &  8.90 $\pm$  0.02 &  0.59 $\pm$  0.03 &  1.74 $\pm$  0.01\\
HD\,91276$^c$        &  44.29 $\pm$   0.63 & 238.94 $\pm$  0.13 &  2.56 $\pm$  0.37 &  7.79 $\pm$  0.00  & 11.43 $\pm$  0.05 &  0.19 $\pm$  0.00 &  0.81 $\pm$  0.01\\
BD\,+13\,2383$^c$    &   1.91 $\pm$   0.45 & 235.69 $\pm$  0.07 & -9.24 $\pm$  0.15 &  8.19 $\pm$  0.01  &  9.92 $\pm$  0.02 &  0.10 $\pm$  0.00 &  0.88 $\pm$  0.02\\
BD\,--07\,3523       &  98.45 $\pm$   5.64 & 65.21 $\pm$  4.10 & 46.18 $\pm$  1.06 &  1.32 $\pm$  0.10  &  8.64 $\pm$  0.10 &  0.74 $\pm$  0.02 &  2.22 $\pm$  0.05\\
HD\,115575$^a$       &  -41.75 $\pm$   2.61 & -100.62 $\pm$  7.76 & 20.72 $\pm$  4.20 &  2.20 $\pm$  0.21  &  7.61 $\pm$  0.03 &  0.55 $\pm$  0.03 &  1.35 $\pm$  0.03\\
BD\,+48\,2167        &  -102.66 $\pm$   2.48 & 32.19 $\pm$  3.42 & -11.10 $\pm$  1.89 &  0.65 $\pm$  0.07  &  9.42 $\pm$  0.06 &  0.87 $\pm$  0.01 &  2.18 $\pm$  0.09\\
BD\,+06\,2880        &  -349.70 $\pm$  12.25 & -13.19 $\pm$  9.67 & -165.05 $\pm$  7.93 &  0.19 $\pm$  0.12  & 49.77 $\pm$ 10.63 &  0.99 $\pm$  0.00 & 21.83 $\pm$  5.15\\
BD\,+32\,2483        &  -84.34 $\pm$   2.61 & 25.28 $\pm$  5.70 & 61.27 $\pm$  1.42 &  0.61 $\pm$  0.14  &  8.36 $\pm$  0.04 &  0.86 $\pm$  0.03 &  4.74 $\pm$  0.28\\
BD\,+41\,2520        &  45.10 $\pm$   0.65 & 63.42 $\pm$  2.27 & 108.43 $\pm$  1.17 &  1.68 $\pm$  0.05  &  8.44 $\pm$  0.02 &  0.67 $\pm$  0.01 &  4.61 $\pm$  0.11\\
HD\,130971$^c$       &  -39.22 $\pm$   0.22 & 197.81 $\pm$  0.63 &  8.35 $\pm$  0.30 &  5.71 $\pm$  0.04  &  7.97 $\pm$  0.02 &  0.16 $\pm$  0.00 &  0.65 $\pm$  0.01\\
BD\,+24\,2817$^c$    &  54.91 $\pm$   0.92 & 197.72 $\pm$  0.60 & -3.98 $\pm$  1.57 &  5.73 $\pm$  0.04  &  8.75 $\pm$  0.01 &  0.21 $\pm$  0.00 &  0.61 $\pm$  0.01\\
HD\,238439$^b$       &  191.91 $\pm$   6.52 & -50.25 $\pm$ 12.20 & 116.43 $\pm$  6.96 &  1.10 $\pm$  0.29  & 15.41 $\pm$  1.16 &  0.87 $\pm$  0.02 &  6.86 $\pm$  0.76\\
HD\,138934$^c$       &  -29.60 $\pm$   0.27 & 232.14 $\pm$  0.23 & 17.50 $\pm$  0.13 &  7.49 $\pm$  0.01  &  9.50 $\pm$  0.01 &  0.12 $\pm$  0.00 &  0.41 $\pm$  0.00\\
HD\,139423           &  -345.55 $\pm$   6.84 & -88.84 $\pm$ 10.14 & 76.66 $\pm$  2.39 &  1.50 $\pm$  0.15  & 36.97 $\pm$  3.53 &  0.92 $\pm$  0.00 & 17.66 $\pm$  1.43\\
HD\,142614$^b$       &  331.80 $\pm$   2.36 & -31.19 $\pm$  4.77 & -51.67 $\pm$  4.14 &  0.47 $\pm$  0.07  & 26.85 $\pm$  0.47 &  0.97 $\pm$  0.00 & 11.05 $\pm$  0.12\\
HD\,143348           &  54.52 $\pm$   0.90 & 87.65 $\pm$  3.21 & 38.02 $\pm$  2.57 &  1.86 $\pm$  0.08  &  8.02 $\pm$  0.00 &  0.62 $\pm$  0.01 &  1.19 $\pm$  0.08\\
BD\,+11\,2896        &  122.65 $\pm$   0.52 &  0.32 $\pm$  3.79 & -87.24 $\pm$  1.19 &  0.01 $\pm$  0.04  &  8.62 $\pm$  0.04 &  1.00 $\pm$  0.01 &  4.32 $\pm$  0.06\\
BD\,+20\,3298$^b$    &  335.50 $\pm$   5.25 & -22.99 $\pm$  5.07 & 98.30 $\pm$  7.70 &  0.30 $\pm$  0.07  & 30.87 $\pm$  2.28 &  0.98 $\pm$  0.00 &  7.04 $\pm$  1.06\\
TYC\,2588-1386-1     &   4.08 $\pm$  12.41 & -48.65 $\pm$  3.54 & -83.86 $\pm$  6.18 &  1.96 $\pm$  0.03  &  8.80 $\pm$  0.39 &  0.64 $\pm$  0.02 &  6.90 $\pm$  0.07\\
TYC\,3085-119-1      &  -22.21 $\pm$   0.51 & 139.98 $\pm$  0.09 & -62.56 $\pm$  0.12 &  3.71 $\pm$  0.00  &  7.91 $\pm$  0.00 &  0.36 $\pm$  0.00 &  1.68 $\pm$  0.01\\
BD\,+39\,3309$^b$    &  207.09 $\pm$   1.80 &  7.16 $\pm$  0.90 & -11.00 $\pm$  1.30 &  0.11 $\pm$  0.01  & 12.16 $\pm$  0.10 &  0.98 $\pm$  0.00 &  1.16 $\pm$  0.01\\
HD\,165400$^c$       &  -11.43 $\pm$   0.11 & 217.20 $\pm$  0.31 & -4.44 $\pm$  0.28 &  7.01 $\pm$  0.03  &  7.64 $\pm$  0.02 &  0.04 $\pm$  0.00 &  0.21 $\pm$  0.00\\
TYC\,1008-1200-1$^b$ &  365.43 $\pm$   3.42 & -28.81 $\pm$  4.85 & -27.82 $\pm$  3.16 &  0.33 $\pm$  0.05  & 29.71 $\pm$  0.60 &  0.98 $\pm$  0.00 &  5.52 $\pm$  0.19\\
TYC\,2113-471-1      &  74.56 $\pm$   2.29 & -27.53 $\pm$  1.28 & -73.78 $\pm$  1.30 &  0.48 $\pm$  0.03  &  7.29 $\pm$  0.05 &  0.88 $\pm$  0.01 &  2.22 $\pm$  0.06\\
TYC\,4221-640-1$^b$  &  -195.33 $\pm$   8.35 &  4.23 $\pm$  4.92 & 81.33 $\pm$  9.43 &  0.14 $\pm$  0.54  & 15.33 $\pm$  1.15 &  0.98 $\pm$  0.05 & 13.93 $\pm$  2.40\\
TYC\,4584-784-1      &  -100.72 $\pm$   2.88 & -38.07 $\pm$  3.71 & -15.45 $\pm$  4.24 &  0.86 $\pm$  0.09  & 11.01 $\pm$  0.14 &  0.85 $\pm$  0.01 &  1.77 $\pm$  0.08\\
TYC\,3944-698-1      &  -83.95 $\pm$   2.19 & 11.73 $\pm$  1.87 & -23.89 $\pm$  1.16 &  0.22 $\pm$  0.04  &  9.54 $\pm$  0.10 &  0.95 $\pm$  0.01 &  1.06 $\pm$  0.01\\
HD\,354750           &  240.06 $\pm$   6.94 & 58.46 $\pm$  6.18 & 41.90 $\pm$  0.41 &  0.87 $\pm$  0.10  & 12.61 $\pm$  0.41 &  0.87 $\pm$  0.02 &  2.51 $\pm$  0.21\\
BD\,--00\,3963$^c$   &  31.86 $\pm$   0.17 & 205.68 $\pm$  0.05 & 17.30 $\pm$  0.06 &  6.25 $\pm$  0.01  &  8.08 $\pm$  0.00 &  0.13 $\pm$  0.00 &  0.32 $\pm$  0.00\\
BD\,+07\,4625        &  -29.08 $\pm$   3.55 & -278.29 $\pm$  1.54 & 248.46 $\pm$  0.39 &  7.57 $\pm$  0.02  & 40.41 $\pm$  0.86 &  0.68 $\pm$  0.01 & 24.78 $\pm$  0.47\\
BD\,+25\,4520        &  150.05 $\pm$   6.23 & 178.22 $\pm$  4.84 & -223.04 $\pm$  9.51 &  5.60 $\pm$  0.06  & 23.87 $\pm$  1.84 &  0.62 $\pm$  0.03 & 16.57 $\pm$  1.94\\
HD\,208316           &  -58.60 $\pm$   4.63 & -58.91 $\pm$  6.55 & -12.86 $\pm$  4.13 &  1.14 $\pm$  0.15  &  7.77 $\pm$  0.04 &  0.74 $\pm$  0.03 &  1.10 $\pm$  0.10\\
TYC\,4267-2023-1$^a$ &  -25.68 $\pm$   1.12 & -132.41 $\pm$  0.77 & 54.88 $\pm$  1.39 &  3.56 $\pm$  0.05  &  8.65 $\pm$  0.01 &  0.42 $\pm$  0.01 &  1.41 $\pm$  0.05\\
TYC\,565-1564-1$^b$  &  236.11 $\pm$  10.06 & -14.20 $\pm$ 12.56 & -17.13 $\pm$  8.83 &  0.24 $\pm$  0.20  & 14.78 $\pm$  1.21 &  0.97 $\pm$  0.02 &  5.27 $\pm$  0.58\\
BD\,+21\,4759        &  -106.37 $\pm$   3.16 & 33.01 $\pm$  0.11 & 42.01 $\pm$  1.76 &  0.63 $\pm$  0.00  &  9.28 $\pm$  0.08 &  0.87 $\pm$  0.00 &  1.53 $\pm$  0.02\\
BD\,+35\,4847        &  109.43 $\pm$   3.64 & 19.10 $\pm$  3.14 & -98.70 $\pm$  4.69 &  0.38 $\pm$  0.06  &  9.82 $\pm$  0.17 &  0.92 $\pm$  0.01 &  4.37 $\pm$  0.44\\
TYC\,2228-838-1$^b$  &  195.91 $\pm$  10.10 &  2.14 $\pm$  9.27 & -3.21 $\pm$  4.35 &  0.04 $\pm$  0.10  & 13.19 $\pm$  0.79 &  0.99 $\pm$  0.01 &  2.37 $\pm$  0.21\\
BD\,--00\,4538       &  164.76 $\pm$   3.70 & -35.84 $\pm$  4.11 & 65.71 $\pm$  2.39 &  0.74 $\pm$  0.09  & 11.05 $\pm$  0.17 &  0.87 $\pm$  0.01 &  4.48 $\pm$  0.06\\
TYC\,4001-1161-1$^b$ &  -305.15 $\pm$   3.81 & -10.34 $\pm$  4.63 & 11.49 $\pm$  1.67 &  0.17 $\pm$  0.07  & 28.67 $\pm$  1.09 &  0.99 $\pm$  0.01 &  0.88 $\pm$  0.04\\
BD\,+03\,4904        &  -179.50 $\pm$   7.26 & 52.61 $\pm$  1.76 & 103.86 $\pm$  3.35 &  1.15 $\pm$  0.07  & 13.90 $\pm$  0.46 &  0.85 $\pm$  0.01 &  5.61 $\pm$  0.18\\
\hline
\multicolumn{8}{l}{(a) Sequoia candidate.}\\
\multicolumn{8}{l}{(b) GSE candidate.}\\
\multicolumn{8}{l}{(c) Thin disc.}\\
\end{tabular}
\tablefoot{Cylindrical galactocentric velocity components (V$_R$, V$_T$, V$_Z$), pericentric (r$_{peri}$) and apocentric (r$_{ap}$) distances, orbit's eccentricities (ecc.) and maximum height over the Galactic plane (Z$_{max}$) obtained using the Galpy code as described in section\,\ref{sec:kin}.}
\end{table*}

\begin{table*}
 \caption{Dynamical properties of the sample.}

\label{kinematics2}
\centering
\begin{tabular}{lcccc}
\hline
Star & E & L$_Z$ & J$_R$ & J$_Z$\\
& km$^2$ s$^{-2}$ & kpc km/s & kpc km/s & kpc km/s\\
\hline
 TYC\,4-369-1$^b$     &  -34227.63 $\pm$  3291.84 & -44.03 $\pm$ 158.21 & 1197.95 $\pm$ 78.51 &  119.72 $\pm$  9.06\\
BD\,+04\,18$^b$      &  -42513.31 $\pm$  1635.62 & 92.87 $\pm$ 75.73 & 1016.43 $\pm$ 71.04 &  45.15 $\pm$  2.83\\
TYC\,33-446-1        &  -50421.65 $\pm$  152.82 & 446.14 $\pm$ 82.86 & 490.18 $\pm$ 46.82 &  179.34 $\pm$  5.00\\
TYC\,2824-1963-1$^b$ &  -32798.81 $\pm$  1356.66 & 464.84 $\pm$ 118.00 & 1047.38 $\pm$ 105.01 &  92.10 $\pm$ 11.10\\
TYC\,4331-136-1      &  -53172.43 $\pm$  407.65 & 341.14 $\pm$ 41.80 & 546.21 $\pm$ 17.06 &  175.52 $\pm$ 23.48\\
HD\,87740$^c$        &  -35668.33 $\pm$  20.63 & 1917.26 $\pm$  0.98 & 67.49 $\pm$  0.92 &   6.29 $\pm$  0.15\\
BD\,+31\,2143$^a$    &  -55728.06 $\pm$  721.22 & -802.73 $\pm$ 58.57 & 301.37 $\pm$ 25.08 &  45.69 $\pm$  0.46\\
HD\,91276$^c$        &  -34291.52 $\pm$  94.96 & 1994.38 $\pm$  2.18 & 49.31 $\pm$  1.25 &  10.46 $\pm$  0.24\\
BD\,+13\,2383$^c$    &  -36851.45 $\pm$  67.52 & 1924.87 $\pm$  1.44 & 11.86 $\pm$  0.22 &  14.82 $\pm$  0.64\\
BD\,--07\,3523       &  -59178.43 $\pm$  310.18 & 477.42 $\pm$ 31.41 & 419.23 $\pm$ 22.81 &  65.91 $\pm$  1.22\\
HD\,115575$^a$       &  -62436.30 $\pm$  569.91 & -740.81 $\pm$ 55.13 & 235.95 $\pm$ 25.27 &  37.22 $\pm$  1.08\\
BD\,+48\,2167        &  -56753.65 $\pm$  246.36 & 261.40 $\pm$ 27.62 & 598.58 $\pm$ 16.69 &  56.42 $\pm$  2.97\\
BD\,+06\,2880        &  5622.52 $\pm$  5730.94 & -92.51 $\pm$ 67.22 & 3584.74 $\pm$ 667.63 &  184.49 $\pm$  8.93\\
BD\,+32\,2483        &  -59630.68 $\pm$  255.87 & 189.39 $\pm$ 43.00 & 443.13 $\pm$ 15.36 &  216.69 $\pm$ 17.41\\
BD\,+41\,2520        &  -57201.08 $\pm$  24.02 & 499.68 $\pm$ 17.95 & 309.05 $\pm$  5.88 &  226.49 $\pm$  7.52\\
HD\,130971$^c$       &  -48991.09 $\pm$  173.44 & 1469.11 $\pm$  7.06 & 28.05 $\pm$  0.68 &  11.49 $\pm$  0.35\\
BD\,+24\,2817$^c$    &  -46355.49 $\pm$  67.06 & 1532.55 $\pm$  4.89 & 47.22 $\pm$  1.59 &   9.15 $\pm$  0.24\\
HD\,238439$^b$       &  -35278.11 $\pm$  3032.28 & -409.69 $\pm$ 100.28 & 920.41 $\pm$ 28.28 &  171.13 $\pm$ 14.78\\
HD\,138934$^c$       &  -39560.78 $\pm$  62.24 & 1828.31 $\pm$  2.59 & 17.45 $\pm$  0.11 &   4.24 $\pm$  0.08\\
HD\,139423           &  -3655.31 $\pm$  3049.07 & -632.91 $\pm$ 69.96 & 2369.28 $\pm$ 212.44 &  232.20 $\pm$  6.38\\
HD\,142614$^b$       &  -14886.95 $\pm$  619.65 & -213.86 $\pm$ 31.79 & 1910.64 $\pm$ 22.81 &  162.79 $\pm$  7.20\\
HD\,143348           &  -61607.48 $\pm$  150.19 & 669.01 $\pm$ 25.39 & 308.18 $\pm$  9.87 &  28.49 $\pm$  2.82\\
BD\,+11\,2896        &  -59067.51 $\pm$  243.19 &  2.27 $\pm$ 27.22 & 599.82 $\pm$  8.54 &  182.99 $\pm$  3.29\\
BD\,+20\,3298$^b$    &  -10201.18 $\pm$  2540.13 & -156.84 $\pm$ 33.80 & 2311.39 $\pm$ 141.13 &  62.76 $\pm$  7.73\\
TYC\,2588-1386-1     &  -54671.06 $\pm$  1380.35 & -339.89 $\pm$ 29.48 & 277.00 $\pm$ 34.50 &  410.02 $\pm$ 38.02\\
TYC\,3085-119-1      &  -55313.29 $\pm$  10.17 & 1085.79 $\pm$  1.01 & 114.68 $\pm$  0.30 &  53.54 $\pm$  0.38\\
BD\,+39\,3309$^b$    &  -46738.94 $\pm$  340.89 & 54.40 $\pm$  6.83 & 977.79 $\pm$ 11.03 &  15.09 $\pm$  0.28\\
HD\,165400$^c$       &  -46100.83 $\pm$  135.27 & 1620.05 $\pm$  4.77 &  2.04 $\pm$  0.08 &   1.64 $\pm$  0.07\\
TYC\,1008-1200-1$^b$ &  -11569.16 $\pm$  711.06 & -173.10 $\pm$ 26.17 & 2235.15 $\pm$ 29.20 &  46.91 $\pm$  0.92\\
TYC\,2113-471-1      &  -67655.11 $\pm$  154.57 & -187.60 $\pm$  8.12 & 452.52 $\pm$ 10.74 &  82.13 $\pm$  3.89\\
TYC\,4221-640-1$^b$  &  -34552.60 $\pm$  3168.92 & 39.58 $\pm$ 47.04 & 905.59 $\pm$ 31.04 &  557.89 $\pm$ 175.20\\
TYC\,4584-784-1      &  -50164.69 $\pm$  653.59 & -366.30 $\pm$ 36.51 & 696.68 $\pm$ 10.85 &  33.24 $\pm$  1.78\\
TYC\,3944-698-1      &  -57398.19 $\pm$  497.16 & 102.71 $\pm$ 17.00 & 728.08 $\pm$  0.83 &  18.32 $\pm$  0.15\\
HD\,354750           &  -44243.32 $\pm$  1272.11 & 379.15 $\pm$ 41.22 & 813.65 $\pm$ 51.11 &  48.93 $\pm$  4.02\\
BD\,--00\,3963$^c$   &  -47007.38 $\pm$  19.34 & 1565.26 $\pm$  0.80 & 17.55 $\pm$  0.11 &   3.27 $\pm$  0.02\\
BD\,+07\,4625        &  1544.83 $\pm$  599.14 & -2125.03 $\pm$ 10.56 & 1481.70 $\pm$ 53.38 &  698.28 $\pm$  2.18\\
BD\,+25\,4520        &  -15380.65 $\pm$  2293.01 & 1376.01 $\pm$ 37.05 & 759.95 $\pm$ 112.46 &  672.75 $\pm$ 63.24\\
HD\,208316           &  -65109.88 $\pm$  682.23 & -437.14 $\pm$ 47.53 & 399.44 $\pm$ 23.29 &  25.56 $\pm$  3.68\\
TYC\,4267-2023-1$^a$ &  -52771.29 $\pm$  211.40 & -1128.09 $\pm$  8.00 & 161.60 $\pm$  3.79 &  36.50 $\pm$  2.29\\
TYC\,565-1564-1$^b$  &  -37633.12 $\pm$  3170.28 & -104.55 $\pm$ 92.33 & 1071.44 $\pm$ 47.87 &  115.58 $\pm$  6.37\\
BD\,+21\,4759        &  -57871.41 $\pm$  351.37 & 265.05 $\pm$  0.98 & 608.16 $\pm$  7.09 &  33.60 $\pm$  1.12\\
BD\,+35\,4847        &  -53688.33 $\pm$  889.31 & 157.10 $\pm$ 25.69 & 644.01 $\pm$ 15.53 &  165.57 $\pm$ 22.46\\
TYC\,2228-838-1$^b$  &  -42910.75 $\pm$  2446.65 & 18.04 $\pm$ 78.59 & 1053.15 $\pm$ 62.93 &  40.17 $\pm$  1.85\\
BD\,--00\,4538       &  -48836.38 $\pm$  650.68 & -286.84 $\pm$ 32.93 & 670.27 $\pm$  1.88 &  141.76 $\pm$  5.30\\
TYC\,4001-1161-1$^b$ &  -12998.56 $\pm$  1348.15 & -95.09 $\pm$ 42.13 & 2248.08 $\pm$ 101.24 &   2.58 $\pm$  0.06\\
BD\,+03\,4904        &  -39338.09 $\pm$  1127.74 & 434.30 $\pm$ 13.76 & 809.03 $\pm$ 52.92 &  145.72 $\pm$ 15.00\\
\hline
\multicolumn{5}{l}{(a) Sequoia candidate.}\\
\multicolumn{5}{l}{(b) GSE candidate.}\\
\multicolumn{5}{l}{(c) Thin disc.}\\
\end{tabular}
 \tablefoot{Stellar orbital energies (E), vertical component of the angular momentum (L$_Z$), and radial and vertical actions (J$_R$, J$_Z$).} 
\end{table*}

\section{Abundances}\label{secabbo}

With \mygi,\, we derived the abundances up to zinc. An example of the data provided is available in the Tables \ref{table_abbA}-\ref{table_abbC}\footnote{Abundances and the linelist adopted for each star are available from CDS \url{http://cdsweb.u-strasbg.fr/cgi-bin/qcat?J/A+A/}}.
 As the uncertainty, we assumed the line-to-line scatter ($\sigma$). In the case that an abundance is based on a single line, we assumed the largest $\sigma$.
Moreover, to these errors, one should add in quadrature the error generated from the assumed stellar atmospheres. 
 Typical errors due to uncertainties in atmospheric parameters are reported in Table\,\ref{tab:errors}. Similar uncertainties were also obtained by \citet[][seen in particular  in their Table\,8,]{Aroa} where two stars with parameters in the range of the MINCE targets were analyzed with the same methods.
\begin{table}
\centering
\caption{Sensitivity of abundances on atmospheric parameters}
\begin{tabular}{l|rrr}
\hline
Element      & $\Delta{\rm T}_{\rm eff}$ & $\Delta\log{\rm g}$ & $\Delta\xi$ \\
             & 100\,K                  & 0.2\,dex       & 0.2\,\kms \\
\hline 
\ion{O}{i}   & 0.03 & 0.12 & 0.01 \\
\ion{Na}{i}  & 0.08 & 0.02 & 0.01 \\
\ion{Mg}{i}  & 0.08 & 0.03 & 0.03 \\
\ion{Al}{i}  & 0.06 & 0.01 & 0.00 \\
\ion{Si}{i}  & 0.03 & 0.02 & 0.01 \\
\ion{S}{i}   & 0.12 & 0.08 & 0.04 \\
\ion{Ca}{i}  & 0.10 & 0.02 & 0.04 \\
\ion{Sc}{ii} & 0.02 & 0.11 & 0.04 \\
\ion{Ti}{i}  & 0.16 & 0.02 & 0.03 \\ 
\ion{Ti}{ii} & 0.01 & 0.08 & 0.06 \\ 
\ion{V}{i}   & 0.18 & 0.01 & 0.00 \\
\ion{Cr}{i}  & 0.14 & 0.02 & 0.03 \\
\ion{Cr}{ii} & 0.04 & 0.08 & 0.01 \\
\ion{Mn}{i}  & 0.14 & 0.02 & 0.01 \\
\ion{Fe}{i}  & 0.12 & 0.01 & 0.05 \\
\ion{Fe}{ii} & 0.08 & 0.11 & 0.05 \\
\ion{Co}{i}  & 0.14 & 0.01 & 0.00 \\
\ion{Ni}{i}  & 0.10 & 0.01 & 0.02 \\
\ion{Cu}{i}  & 0.13 & 0.01 & 0.00 \\
\ion{Zn}{i}  & 0.04 & 0.06 & 0.03 \\
\hline                           
\end{tabular}                    
\label{tab:errors}             
\end{table}  

When not specified, we adopted  the abundance derived from \ion{Fe}{i} lines as the metallicity.
Since our surface gravities are derived form the parallaxes
and not Fe ionisation equilibrium, in order to minimise the gravity
dependence in abundance ratios, we adopted [X/Fe] =  [X/\ion{Fe}{i}], where
X is a neutral species and [X/Fe] = [X/\ion{Fe}{ii}] for ionised species.
The exception is oxygen: since all our oxygen abundances are derived from the forbidden
lines, whose dependence on surface gravity is closer to that of an ionised species
than to that of a neutral species we adopt [O/Fe] = [O/\ion{Fe}{ii}] as done by \citet{Cayrel04}.
The solar abundances we adopted are reported in Table\,\ref{tab:solarabbo} and these are
the values we used to computed our stellar models as well as to derive [X/H] and [X/Fe] ratios.

 A slightly different approach was adopted to derive the abundances of sulphur. Both the S I lines of Multiplet 1 at 920\,nm and Multiplet 6 at 860\,nm lie in the wavelengths ranges covered only by the spectra taken at CFHT with the spectrograph ESPaDOnS, but we considered only the lines at 920\,nm because those at 860\,nm are too weak to be measured.
The strong \ion{S}{i} lines of Multiplet 1 are located in a wavelength range contaminated by telluric absorptions.
To assess the suitability of Mult. 1 lines for the estimation of sulphur abundances, we compared the observed spectrum of our stars with that of a B-type star.
Sulphur lines contaminated by telluric ones were rejected, while we derived the sulphur abundances from not contaminated lines by spectrosynthesis. We used the code SALVADOR (Mucciarelli in prep.) that computes a grid of synthetic spectra with the code SYNTHE    \citep{Kurucz1993b, Kurucz2005}, using ATLAS9 $\alpha$-enhanced model atmospheres \citep{Kurucz1993a} based on ODF by \cite{Castelli_Kurucz2003}. This code allows us to speed up the fitting procedure and it keeps the results consistent with the other elements because based on the same codes to compute theoretical synthesis and models atmospheres.
SALVADOR  (in the same way as MyGIsFOS) finds the abundance from a line performing a $\chi^2$ minimisation between observed and synthetic spectra.
The difference in this case is that in the grid of synthetic spectra used by \mygi, all the abundances scale with Fe, while SALVADOR computes the synthesis where just the S abundance changes.

 In cases where the S lines were contaminated by telluric absorptions, we estimated the abundances from  equivalent width
(EW). In this way, by using the deblending option of IRAF, we could estimate in the feature contaminated by telluric lines the contribution in EW from the telluric line and the one from S.

The measured EW were converted in abundances using the code GALA. As explained by \citet{Mucciarelli2013}, GALA computes the curve of growth of an element by using WIDTH code \citep{kurucz05} and computing an ATLAS model. In this way the S abundance derived from spectralsynthesis and EW are perfectly consistent.
The results obtained are listed in Table \ref{sulphur}. We reported also the abundances corrected for deviations from local thermodynamic equilibrium (LTE). The non-LTE corrections were assumed following \cite{Takeda2005} and we found a mean correction of $\Delta \sim -0.25$.
The [S/Fe] values were obtained considering the solar value [S/Fe]$_\odot$=7.16 \citep{Caffau2014}.  

\begin{table}
\caption{Sulphur abundances.}

\begin{center}
\begin{tabular}{l|c|c|c}
\hline
ID & A(S)$_{LTE}$ & $A(S)_{NLTE}$ & [S/Fe]$_{NLTE}$ \\
\hline
BD$+$39$-$3309   & 5.33 & 5.00 & 0.42 \\
BD$+$31$-$2143  &  5.50 & 5.20 & 0.41 \\
BD$+$32$-$2483  & 5.76  & 5.51 & 0.60 \\
BD$+$20$-$3298  & 5.89  & 5.82 & 0.61 \\
BD$+$48$-$2167  & 5.87  & 5.57 & 0.69 \\
TYC\,3085$-$119$-$1 & 6.05 & 5.79 & 0.14\\
\hline
\end{tabular}
\end{center}
\label{sulphur}
\tablefoot{From lines of Multiplet 1 in spectra obtained at CFHT with ESPaDOnS.}
\end{table}

\begin{table}
\caption{Solar abundances used throughout this paper.}
\label{tab:solarabbo}
\begin{tabular}{lll}
\hline
Element & A(X) & Reference \\
\hline
C  & 8.50 & \citet{Caffau11} \\
O  & 8.76 & \citet{Caffau11} \\
Na & 6.30 & \citet{Lodders09} \\
Mg & 7.54 & \citet{Lodders09} \\
Al & 6.47 & \citet{Lodders09} \\
Si & 7.52 & \citet{Lodders09} \\
S  & 7.16 & \citet{Caffau11} \\
Ca & 6.33 & \citet{Lodders09} \\
Sc & 3.10 & \citet{Lodders09} \\
Ti & 4.90 & \citet{Lodders09} \\
V  & 4.00 & \citet{Lodders09} \\
Cr & 5.64 & \citet{Lodders09} \\
Mn & 5.37 & \citet{Lodders09} \\
Fe & 7.52 & \citet{Caffau11} \\
Co & 4.92 & \citet{Lodders09} \\
Ni & 6.23 & \citet{Lodders09} \\
Cu & 4.21 & \citet{Lodders09} \\
Zn & 4.62 & \citet{Lodders09} \\
\hline
\end{tabular}
\end{table}

\section{Chemical evolution of the MINCE stars}

The main scope of Galactic archaeology is to constrain the formation and evolution of the Milky Way from the observed chemical abundances.
Hence, in Sects. \ref{Aelem}-\ref{Celem},   we  compare  the  stellar abundance ratios from the MINCE project  with the predictions of two chemical evolution models. 
In Sect. \ref{CEM}, we briefly recall the main characteristics of the reference chemical evolution we used in this study for:  i) the Milky disc  and ii)  the
GSE accretion event, respectively.

\subsection{Reference chemical evolution models}\label{CEM}

\subsubsection{Model for the disc components by \citet{Spitoni21} }

The inner halo \citep{Nissen2010} and the oldest stars of the thick disk share the same chemical enrichment and it is likely that they were both formed during the same dissipative collapse process. This is why we decided to compare MINCE data with \citet{Spitoni21}, which provided a reliable model for the Milky disc components constrained by high-resolution spectroscopy data using a Bayesian framework. Moreover, we think that it is important also to show the low-$\alpha$ evolution (thin disc) of \citet{Spitoni21} because the youngest stars predicted by the GSE model (with [Fe/H] =-0.5 dex, see Fig. 6) seem to share the same abundance ratio [X/Fe] (where X=O, Mg, Ca, Si, Ti, Sc , Co, Mn) of the low-metallicity tail of the thin disc phase. 
\citet{Spitoni21} presented a revised version of the classical two-infall chemical evolution  model \citep{chiappini1997} in order to reproduce the Galactic disc components  as traced by the  APOGEE DR16  \citep{Ahumada2020} abundance ratios. In this model, the  Galactic disc is  assumed to be formed by two independent, sequential episodes of gas accretion giving rise to the thick and thin disc components, respectively. 
As already pointed out by \citet{spitoni2019,spitoni2020} and \citet{palla20}   the signature of a delayed gas-rich merger  (i.e. the delay between the two gas infall is   $\sim$ 4 Gyr)  is imprinted in the APOGEE abundance ratios.
We recall that in \citet{Spitoni21} a Bayesian framework based on MCMC methods has been used to find the best  chemical evolution model constrained by  APOGEE DR16  [Mg/Fe] and [Fe/H] abundance ratios at different Galactocentric distances. 
In the solar neighbourhood, the dilution effect caused by the second infall produces a characteristic feature   in the [$\alpha$/Fe] and [Fe/H] space.  In fact, the late accretion of pristine gas has the effect of decreasing the metallicity of stars born immediately after the infall event, leading to  a evolution at nearly constant [$\alpha$/Fe]  since both $\alpha$ and Fe are diluted by the same amount \citep{spitoni2019}.
The \citet{Scalo86} initial stellar mass function (IMF), constant in time and space, was also adopted.

The model computed in the solar vicinity (8\,kpc) assumes primordial infall for both infall episodes but different star formation efficiencies (SFEs):   2 Gyr$^{-1}$ (thick disc)  and  1 Gyr$^{-1}$ (thin disc). We refer the reader to the middle column (model for the solar vicinity) of Table 2 in \citet{Spitoni21} for the values of  the best-fit model parameters as predicted by MCMC calculations: namely, gas infall timescales,   present-day total surface mass density ratio and  the  delay between the two gas infall.
It is worth mentioning that the predicted   present-day total surface mass density ratio   between thin and thick disc sequences of  $5.635^{+0.214}_{-0.162}$  is  in very good agreement  with the value derived by \citet{fuhr2017} for the local mass density ratio (5.26).

\begin{figure}[h]

\includegraphics[width=\hsize,clip=true]{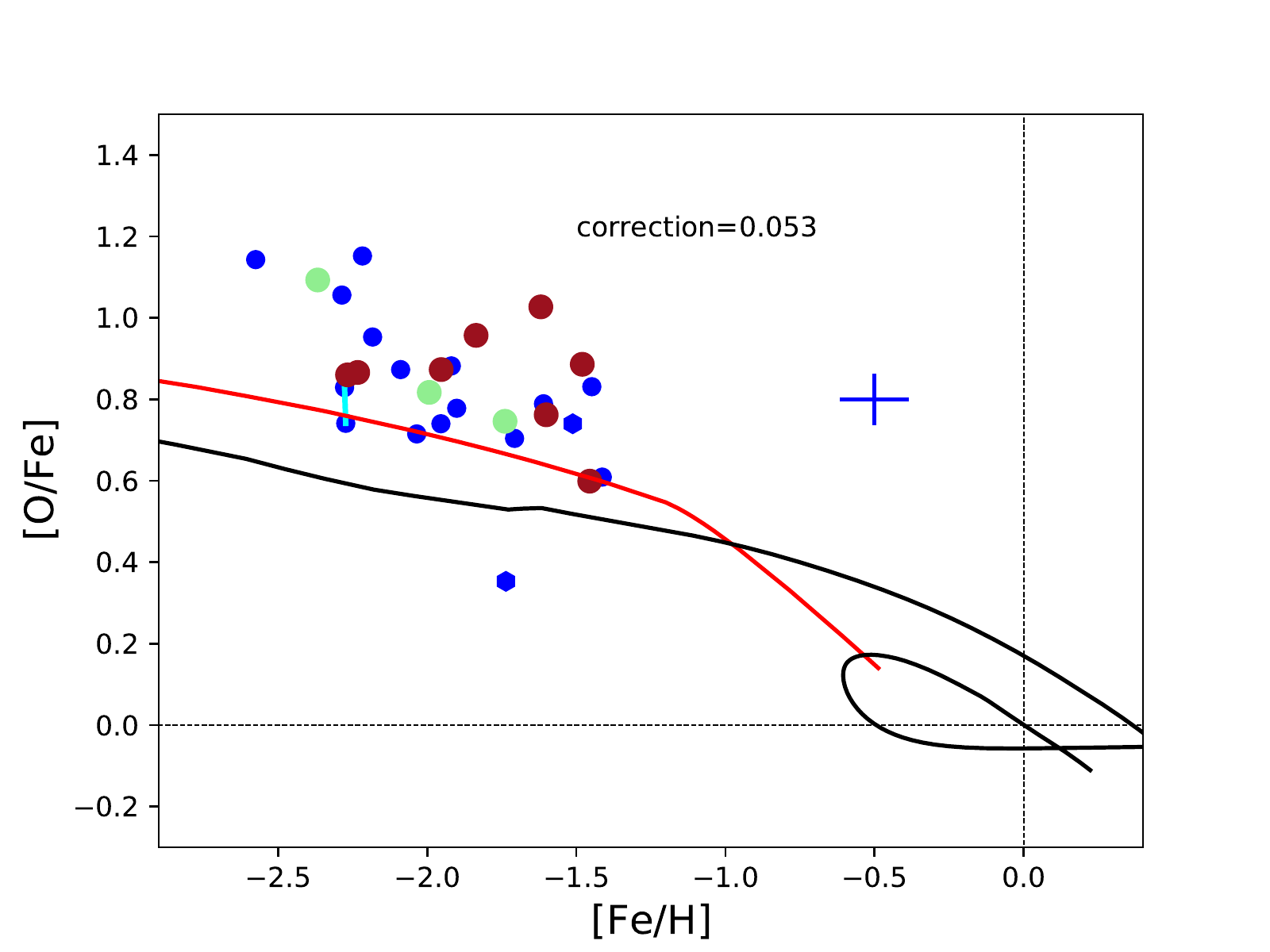}
\caption{[O/Fe] vs [Fe/H] abundances measured in the MINCE stars: blue solid dots stars selected with the mince2 selection, hexagons from APOGEE. Cyan lines connect the abundances measured for the same stars with spectra taken from two facilities. The colours of the dot refer to the substructures to which they belong: red for GSE, light green for Sequoia and blue for the remaining stars.
The mean errors of the MINCE sample are reported as blue cross.
The black lines refers to the chemical evolution model by \citet{Spitoni21} for the discs of our Galaxy, whereas the red line is a model for GSE \citep{Cescutti20}. Correction refers to the offset in [O/Fe] applied to the models (see Sect.\ref{CEM} for details).}
\label{OFe}
\end{figure}

\begin{figure}[h]
\centering
\includegraphics[width=\hsize,clip=true]{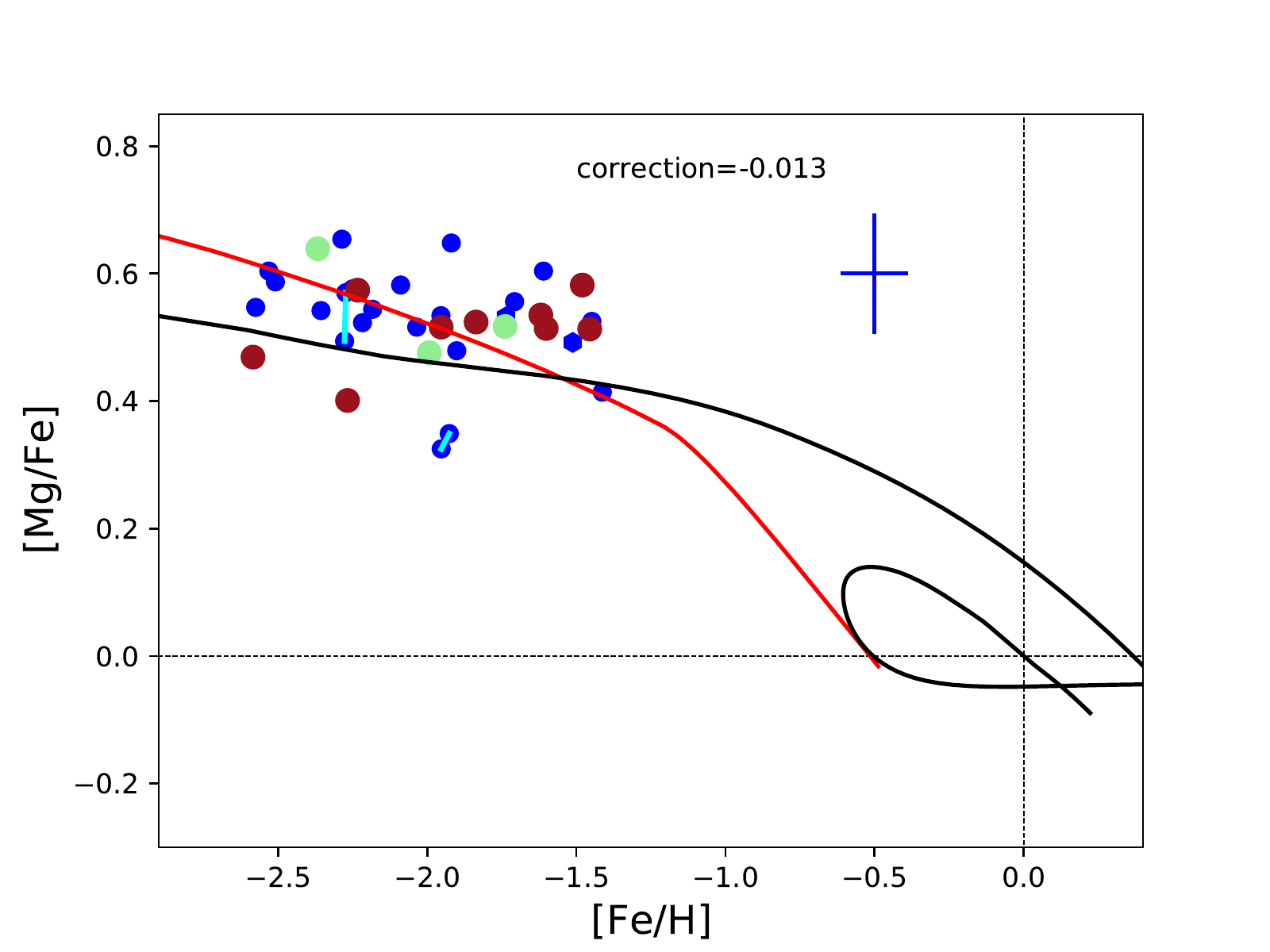}
\caption{[Mg/Fe] vs [Fe/H] abundances measured in the MINCE stars;
the details are the same as Fig.\ref{OFe}.}
\label{MgFe}
\end{figure}

\begin{figure}[h]
\centering
\includegraphics[width=\hsize,clip=true]{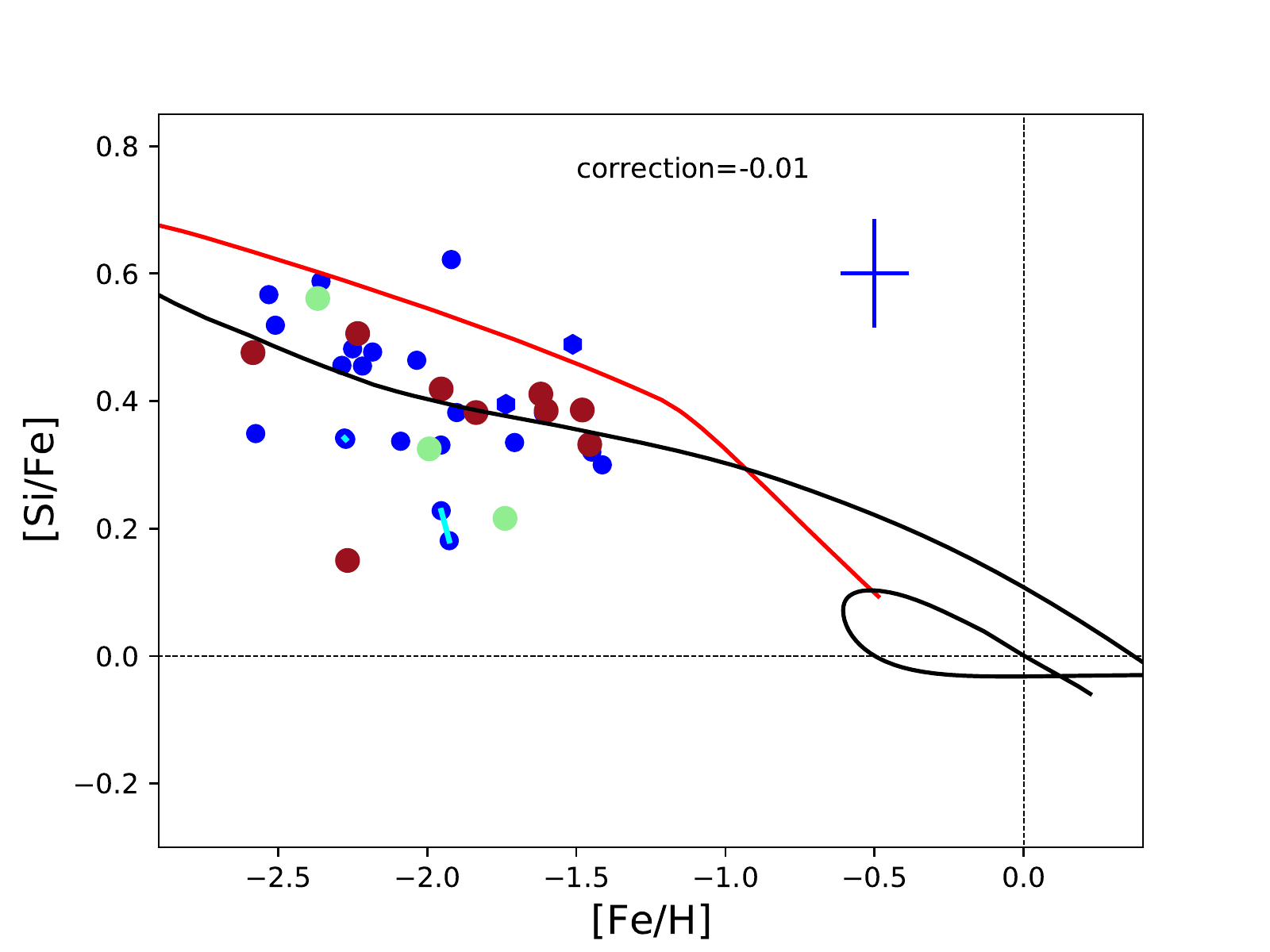}
\caption{[Si/Fe] vs [Fe/H] abundances measured in the MINCE stars; the details are the same as Fig.\ref{OFe}.}
\label{SiFe}
\end{figure}

\begin{figure}[h]
\centering
\includegraphics[width=\hsize,clip=true]{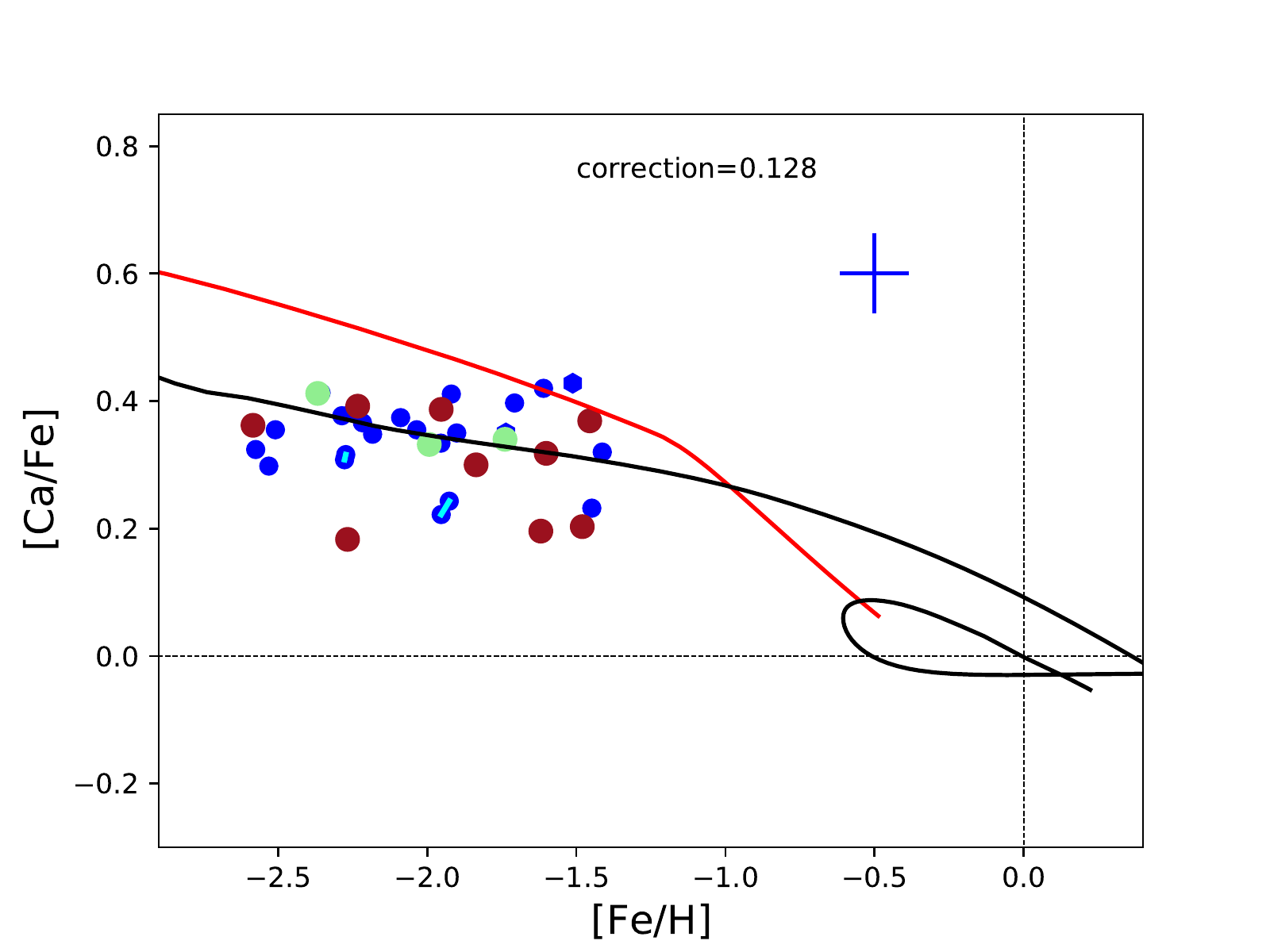}
\caption{[Ca/Fe] vs [Fe/H] abundances measured in the MINCE stars; the details are the same as Fig.\ref{OFe}.
}
\label{CaFe}
\end{figure}

\begin{figure}[h]
\centering
\includegraphics[width=\hsize,clip=true]{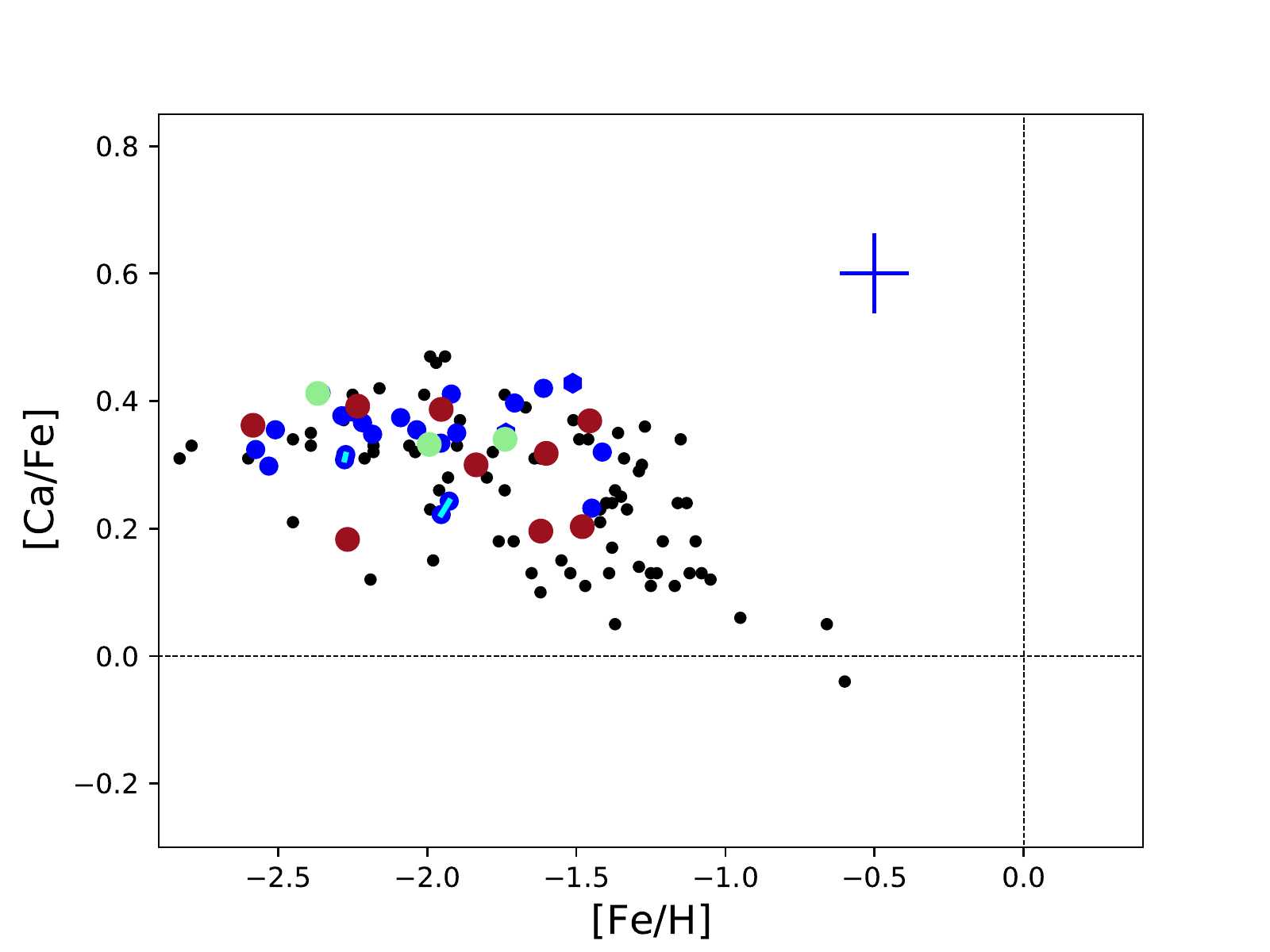}
\caption{[Ca/Fe] vs [Fe/H] abundances measured in the MINCE stars; the plot is the same as Fig.\ref{CaFe}, but without the lines of models and with the results by \citet{ishigaki12} for halo stars (black dots).}
\label{CaFe2}
\end{figure}

\begin{figure}[h]
\centering
\includegraphics[width=\hsize,clip=true]{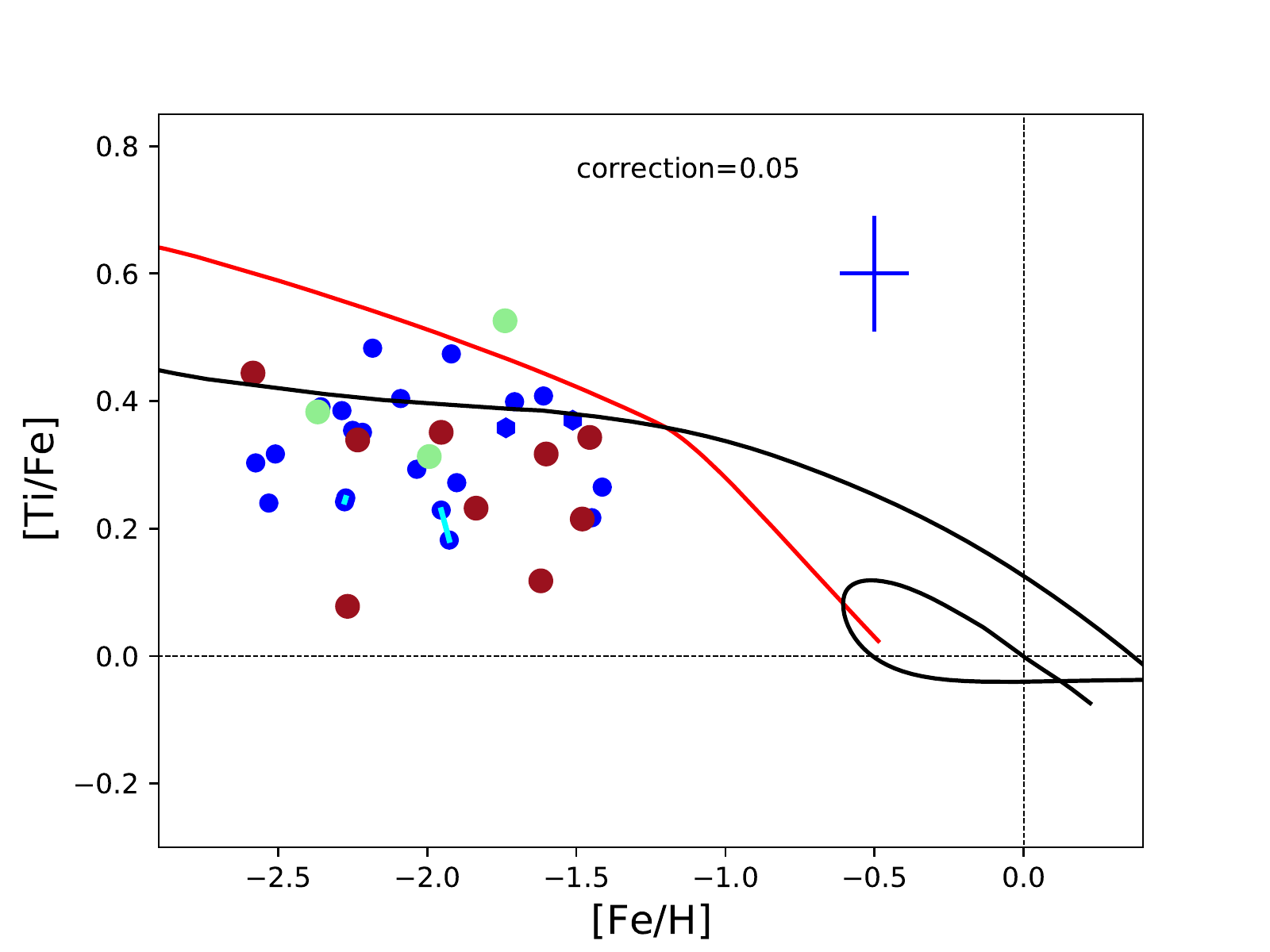}
\caption{[Ti/Fe] vs [Fe/H] abundances measured in the MINCE stars; the details are the same as Fig.\ref{OFe}.}
\label{TiFe}
\end{figure}

\begin{figure}[h]
\centering
\includegraphics[width=\hsize,clip=true]{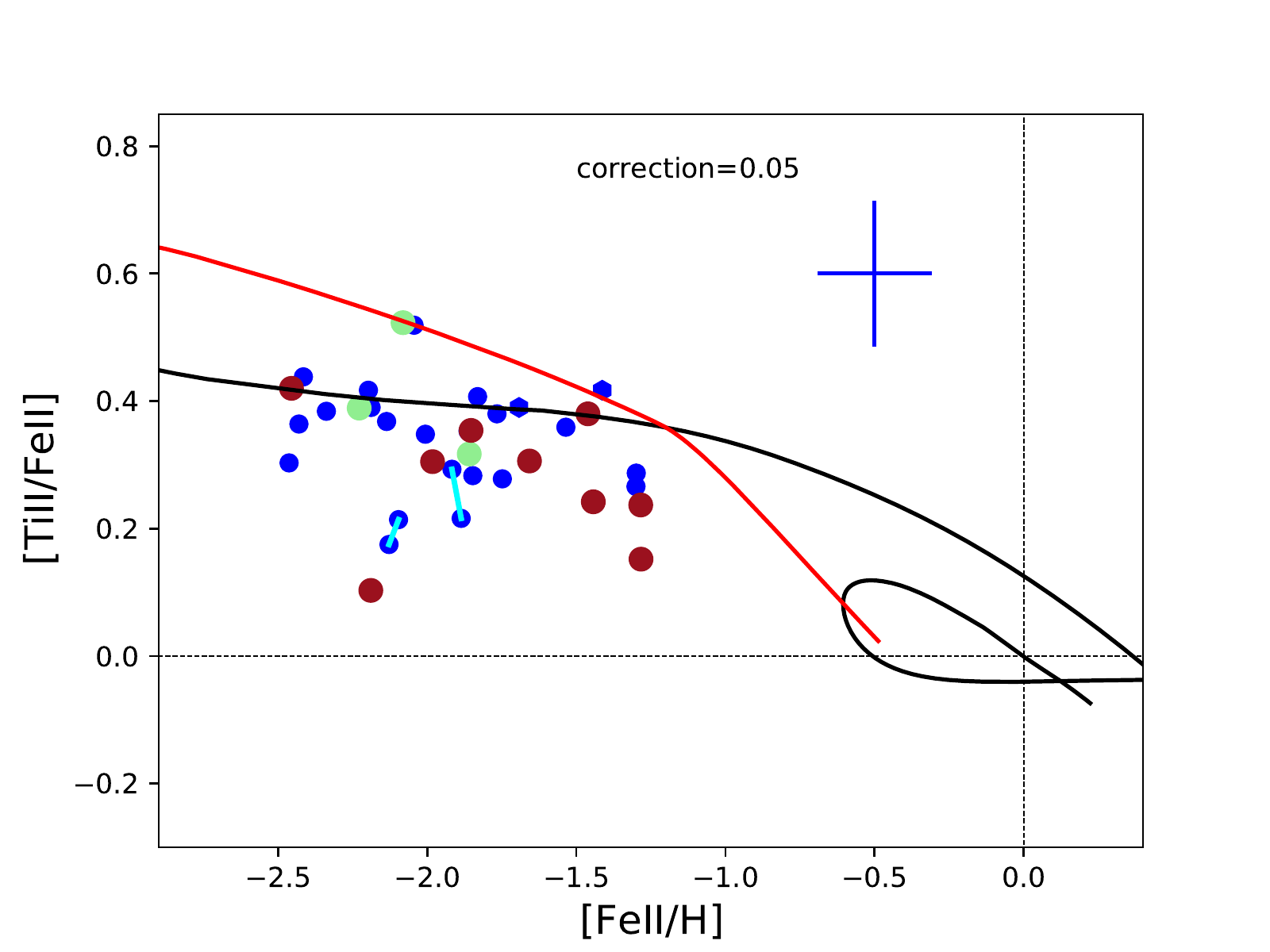}
\caption{[Ti II/Fe II] vs [Fe II/H] abundances measured in the MINCE stars; the details are the same as Fig.\ref{OFe}.}
\label{TiIIFe}
\end{figure}

In this paper, we compare observational data for $\alpha$ and iron-peak  elements with model predictions  in the solar neighbourhood adopting the same  nucleosynthesis prescriptions  as in \citet{Spitoni21}, i.e. applying the ones  suggested by \citet{Franc04}.
This set of yields  has been widely used in the past
\citep{cescutti2007,Mott2013, spitoni2015,
  spitoni2D2018, spitoni2019} and turned out to be able to reproduce the main chemical abundances of the solar neighbourhood.
For most of the elements, the offsets of the model to the solar abundances is very small. However, we decided to apply a correction for each element to have the chemical evolution models passing exactly to [X/Fe]=0 at [Fe/H]=0. This correction is quoted for each element in the relative plot (Fig.s\ref{OFe}-\ref{ZnFe}.
The elements Na, Al, V, Cr, and Cu were not considered in \citet{Franc04} and we do not show model results for these elements here.

\subsubsection{GSE}
A large fraction of the halo stars in the solar vicinity are the result of an accretion event, associated to a disrupted satellite, dubbed GSE \citep{Belokurov18,Haywood18,Helmi18}. As presented in the previous section, we study the kinematics of our sample and we can determine what the progenitors of our sample are, that is, if they used to belong to GSE or Sequoia. For this reason, we compared our data to a model built to describe the chemical enrichment evolution in GSE.
In the following, we summarise the main characteristics of the model. 
The infall law is:
\begin{equation}
A(t)=M_{Enc} Gauss(\sigma_{Enc},\tau_{Enc})
,\end{equation}
where Gauss is a normalised Gaussian function, $\tau_{Enc}$ is time of the center of the peak and $\sigma_{Enc}$ the standard deviation, and $M_{Enc}$ is the total amount of the gas infall into Gaia-Enceladus.
The star formation rate (SFR) is:
\begin{equation}
\psi(t)=
\bigg \{
\begin{array}{rl} 
\nu_{Enc} \Sigma(t)^k & t \leq T_{Enc} \\
0 & t > T_{Enc} \\
\end{array}
,\end{equation}
where $\nu_{Enc}$ is the efficiency of the star formation, $\Sigma(r)$ is the surface mass density, and  the exponent, $k$, is set equal to 1.5 \citep{Kennicutt1989}, $T_{Enc}$ is the time when Gaia-Enceladus stops forming star, due to the interaction with the Galaxy. 
A Galactic wind is considered  as follows:
\begin{equation}
W(t) =
\bigg \{
\begin{array}{rl} 
\nu^{wind}_{Enc}\psi(t) & t \geq T^{wind}_{Enc} \\
 0 & t < T^{wind}_{Enc} \\
\end{array}
,\end{equation}
where $T^{wind}_{Enc}$ is when the galactic wind in Gaia-Enceladus starts due to interaction with the Galaxy and  $\nu^{wind}_{Enc}$ is the wind efficiency.
The seven parameters -- $\nu_{Enc}$, M$_{Enc}$,$\tau_{Enc}$, $\sigma_{Enc}$  T$_{Enc}$, T$^{wind}_{Enc}$, and $\nu^{wind}_{Enc}$ -- determine the equations of the chemical evolution model for Gaia-Enceladus and they are summarised in Table \ref{tab:tab_par}.
The precise procedure is described in \citet{Cescutti20}.  
\begin{table}
 \caption{Parameters for the chemical evolution of Gaia-Enceladus.}
 \label{tab:tab_par}
 \begin{tabular}{l|c}
  \hline
  parameter  & best value after minimisation\\
  \hline
$\nu_{Enc}$ (star formation efficiency) & 1.3 Gyr $^{-1}$\\
M$_{Enc}$ (surface mass density) & 2.0 M$_{\odot}/pc^2$  \\
$\tau_{Enc}$ (peak of the infall law)& 550 Myr \\
$\sigma_{Enc}$ (SD of the infall law)& 1408 Myr \\
$\nu^{wind}_{Enc}$ (galactic wind efficiency) & 5.0 \\
T$^{wind}_{Enc}$ (start of the galactic wind) & 2919 Myr \\
T$_{Enc}$ (end of the star formation) & 5767 Myr  \\
 \hline
 \end{tabular}
\end{table}

To summarise the most important feature, its evolution is similar to a dwarf spheroidal galaxy \citep{LMC08}, namely, it is less massive than our Galaxy by  around a factor of 30 at the beginning. However, given its galactic winds and the less efficient star formation period ending 6 Gyr ago, its stellar content is only a hundredth of the Galactic stellar mass.
The nucleosynthesis adopted is basically the same of  \citet{Franc04}, to be consistent with the model in \citet{Spitoni21}. The only difference is that  the iron  production assumed for SNe II is 0.07\,M$_{\rm \odot}$
 for the SNe II \citep[see]{Limongi18} in \citet{Cescutti20}, which is about a factor of 2 lower than the iron consider in \citet{Franc04}. For this reason,  we decided to decrease accordingly  the yields for iron peak elements from massive stars by a factor of 2; any other deviation is described for the specific element. 

\subsection{Results for $\alpha$-elements }\label{Aelem}

In Figs. \ref{OFe}, \ref{MgFe}, \ref{SiFe}, \ref{CaFe}, \ref{TiFe}, and \ref{TiIIFe}, we show our results for the $\alpha$-elements O, Mg, Si, Ca, and Ti, respectively. 
In these plots, we include only those stars selected with a mince2 selection (and two stars chosen thanks to the APOGEE survey), providing  stars belonging  either to the halo  or to the substructures GSE and Sequoia that we distinguish thanks to the colour coding. 

Searching for specific differences between halo stars and Sequoia or GSE, but also between GSE and Sequoia, we cannot find within $\alpha$-element abundances any clear signal, all the stars seem to share the same plateau with some dispersion. Surely,  the star with the lowest [$\alpha$/Fe] and a relative low [Fe/H] ($\sim -$2.25) belongs to GSE. The other feature that we can be acknowledged is that at [Fe/H]$\sim -1.5$, GSE stars show on average a lower [$\alpha$/Fe]. 

Overall, this outcome is confirmed by the models results. In the models, the chemical differences expected between GSE and the disc of our Galaxy reside in the [$\alpha$/Fe] vs [Fe/H] observed at [Fe/H]$>-$1.5, where the two models do show a clear difference. However, most of our stars have lower metallicity so they are not expected to be firmly distinguished chemically.
Between the two models describing the chemical evolution of the discs of our Galaxy and the GSE, there is also an offset for what concern the plateau, with the one for GSE on average slightly (0.1-0.2) above the  trend of the discs model. This outcome is most likely connected to the choice of the iron yields.  Probably, more interesting is that the models predict a slightly different steepness in the trend of $\alpha$ in the range $-$2.5$<$[Fe/H]$<-$1.5, with GSE having a more negative behaviour. 
A hint towards this can be found in the fact that GSE stars show on average a lower [$\alpha$/Fe] mentioned before, and surely more data may help confirming this feature. 

In Fig. \ref{CaFe2}, we decided to present our results for [Ca/Fe] compared to the results obtained by \citet{ishigaki12} for this element.
The abundances taken from by \citet{ishigaki12} were re-normalised to our solar abundances to avoid spurious offsets. 
Overall, our results are in excellent agreement with the abundances obtained for halo stars by \citet{ishigaki12}, although they can comprise also stars of GSE and Sequoia. The main visible difference is that this sample extends further at metallicity up to [Fe/H]$\sim -$1  (with two stars almost at [Fe/H]$\sim -$0.5).


 
\subsection{Results for sodium and aluminium}\label{Belem}

\begin{figure}[h]
\centering
\includegraphics[width=\hsize,clip=true]{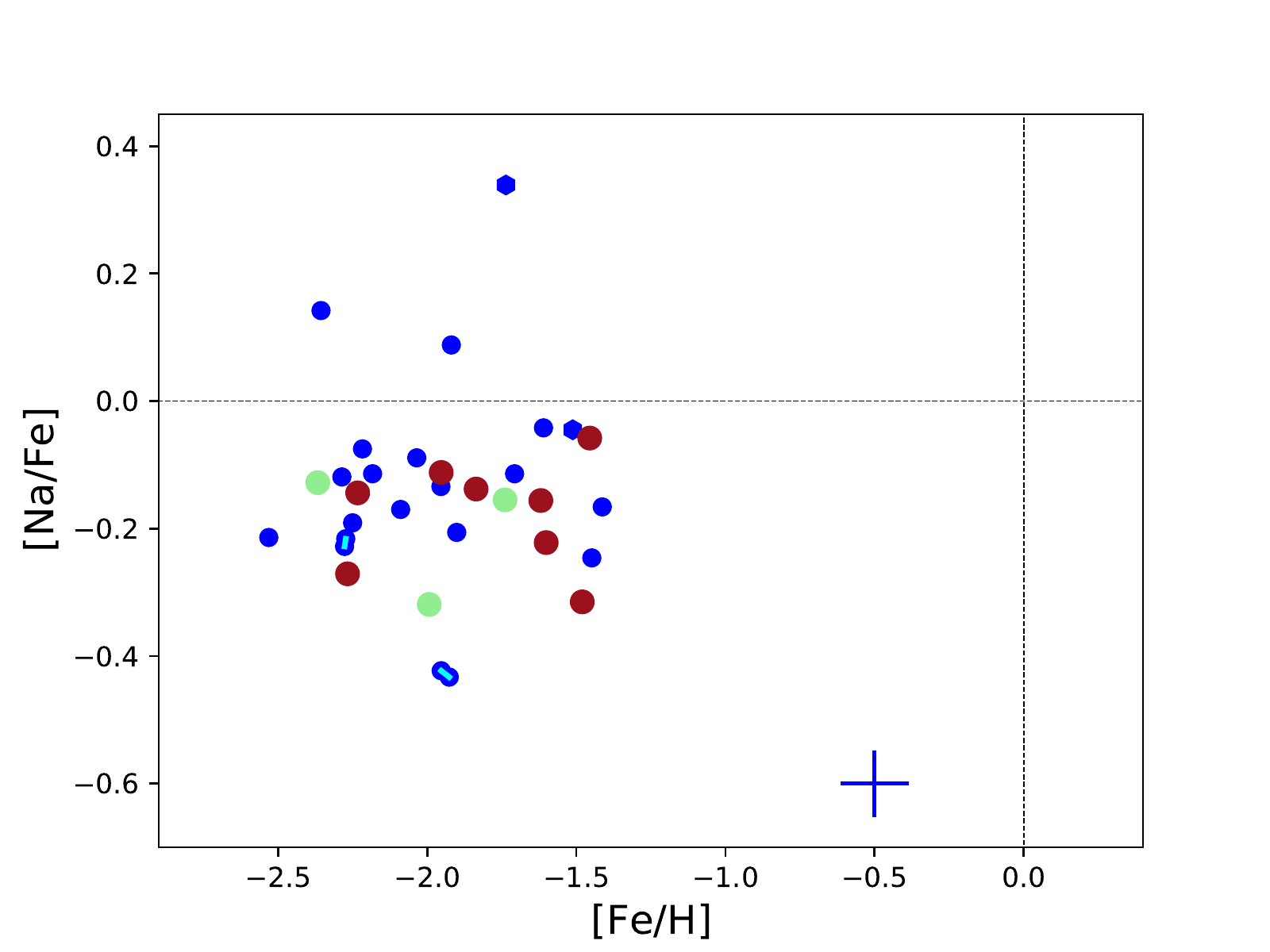}
\caption{[Na/Fe] vs [Fe/H] abundances measured in the MINCE stars; the details are the same as Fig.\ref{OFe},  but without the lines of models.}
\label{NaFe}
\end{figure}

\begin{figure}[h]
\centering
\includegraphics[width=\hsize,clip=true]{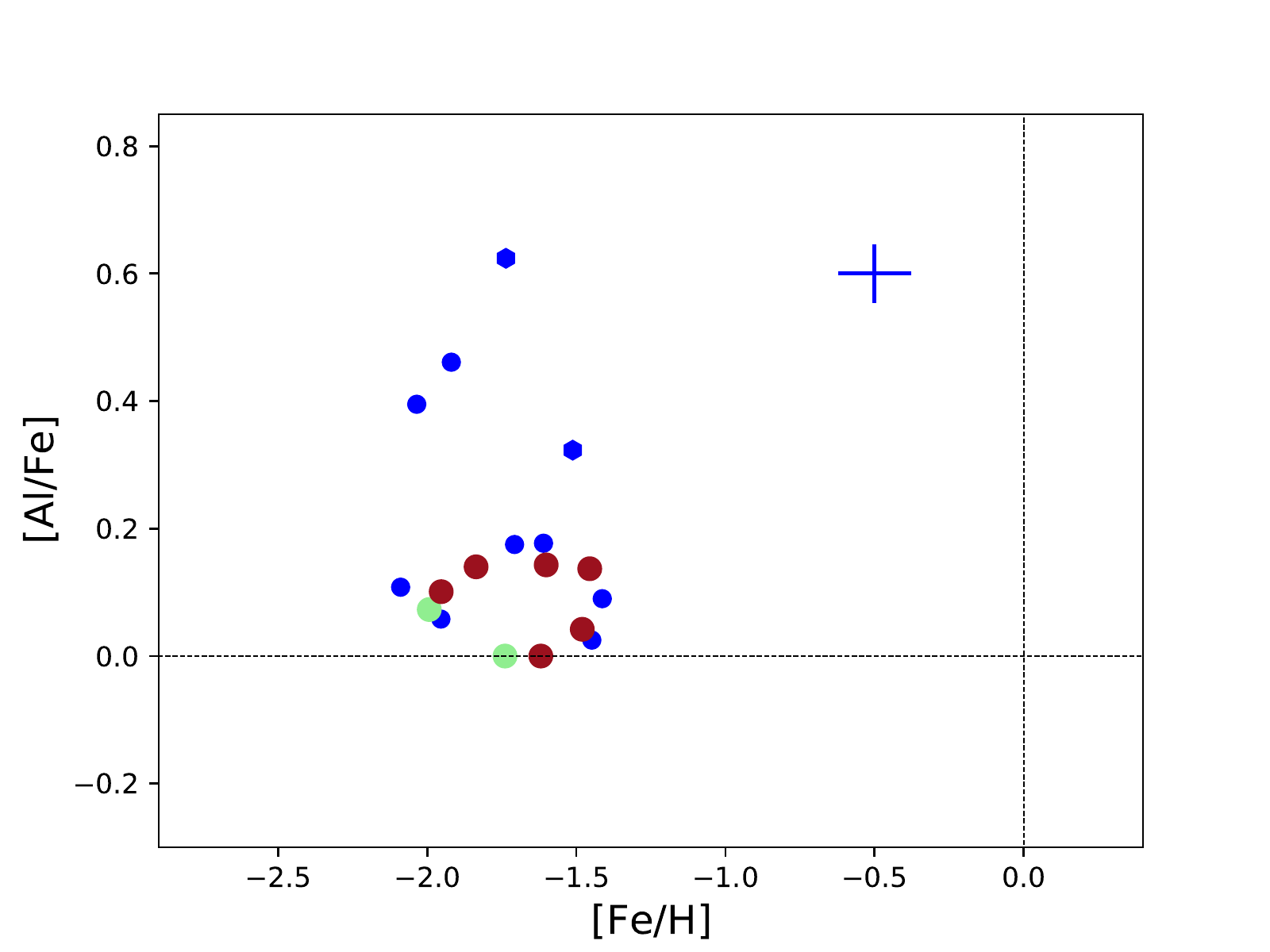}
\caption{[Al/Fe] vs [Fe/H] abundances measured in the MINCE stars; the details are the same as Fig.\ref{OFe},  but without the lines of models.}
\label{AlFe}
\end{figure}

In Fig.\ref{NaFe}, we present the plot for sodium. 
It presents a  dispersion in the data that can be at least partially due to NLTE effects; surely, it is an element where NLTE effects play an important role (see Sect. \ref{nlte} and Table \ref{tab:NLTE}). 
An important feature is visible in the comparison between halo stars and GSE and Sequoia stars. In our sample, we have three stars enhanced in sodium and they all belong to the halo. Given the limited sample, strong conclusions cannot be obtained, but we will follow up this signature
within our future MINCE stars.
We show in Fig. \ref{AlFe} the ratio obtained for aluminium; the dispersion is not present for this element but again four stars show an enhancement of [Al/Fe]$>$0.4 and again none of these stars belong to GSE or Sequoia. 

\subsection{Results for iron-peak elements }\label{Celem}
\begin{figure}[h]
\centering
\includegraphics[width=\hsize,clip=true]{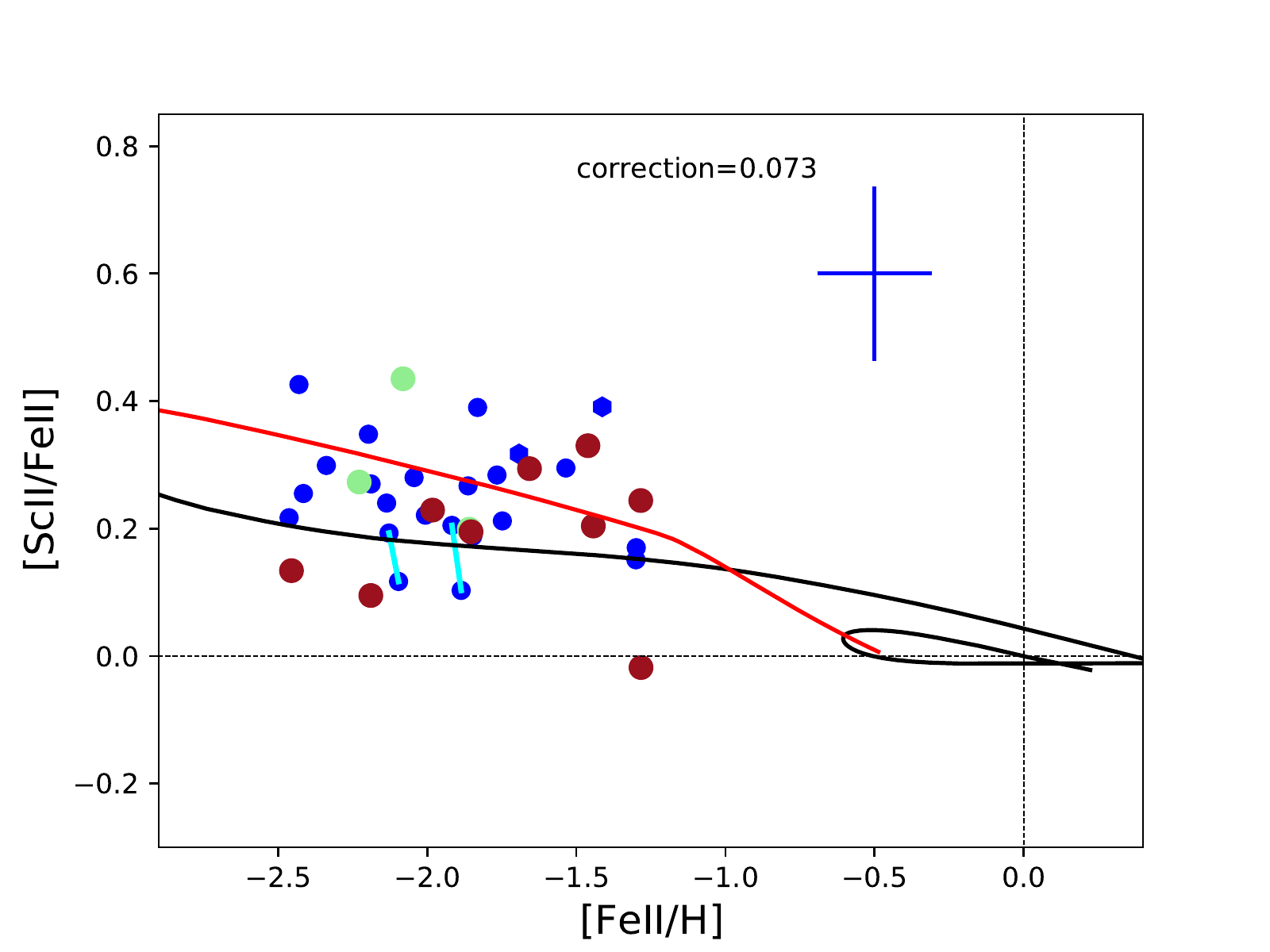}
\caption{[Sc II/Fe II] vs [Fe II/H] abundances measured in the MINCE stars; the details are the same as Fig.\ref{OFe}. }
\label{ScFe}
\end{figure}

\begin{figure}[h]
\centering
\includegraphics[width=\hsize,clip=true]{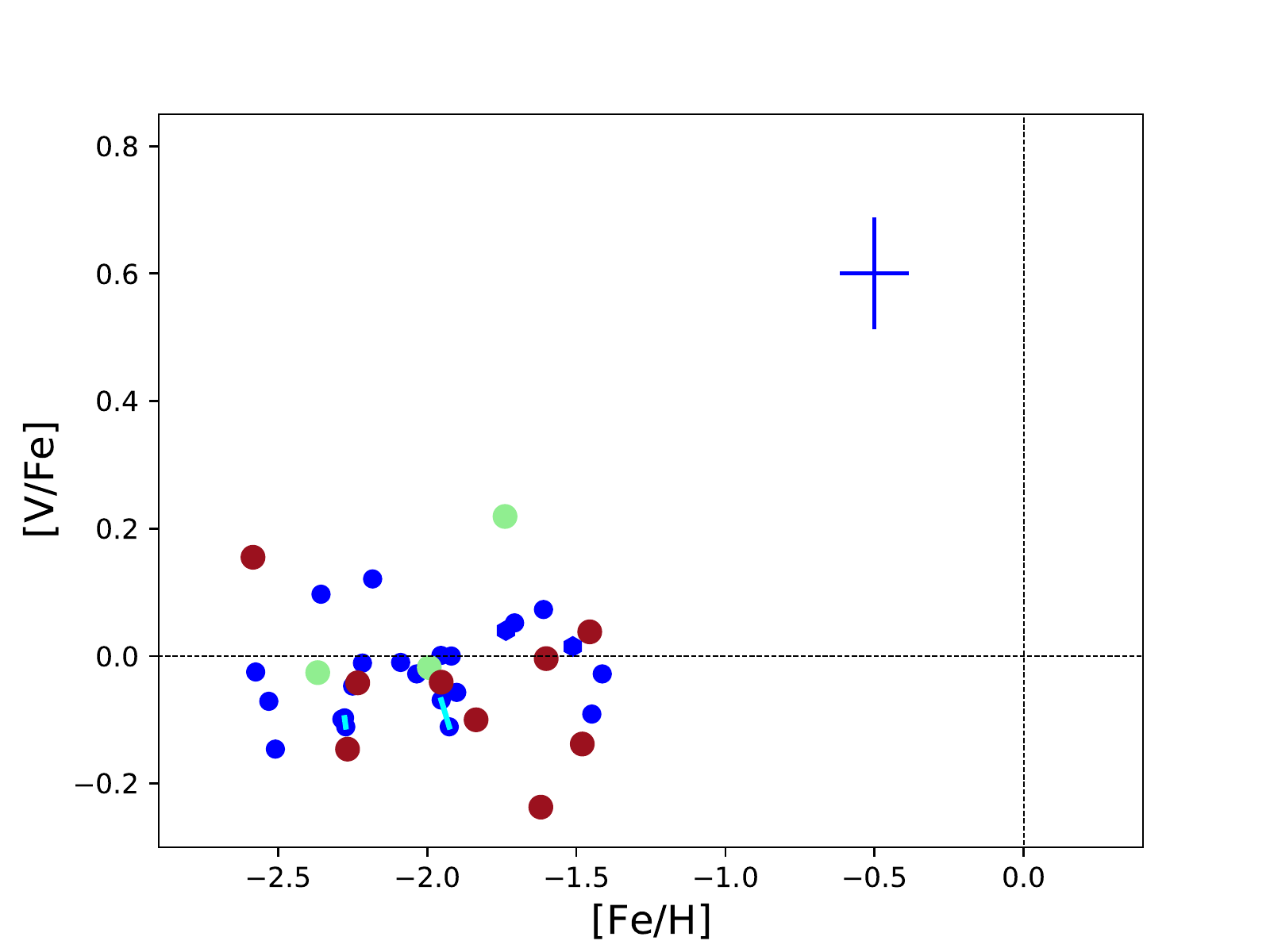}
\caption{[V/Fe] vs [Fe/H] abundances measured in the MINCE stars; the details are the same as Fig.\ref{OFe}, but without the lines of the models. }
\label{VFe}
\end{figure}

\begin{figure}[h]
\centering
\includegraphics[width=\hsize,clip=true]{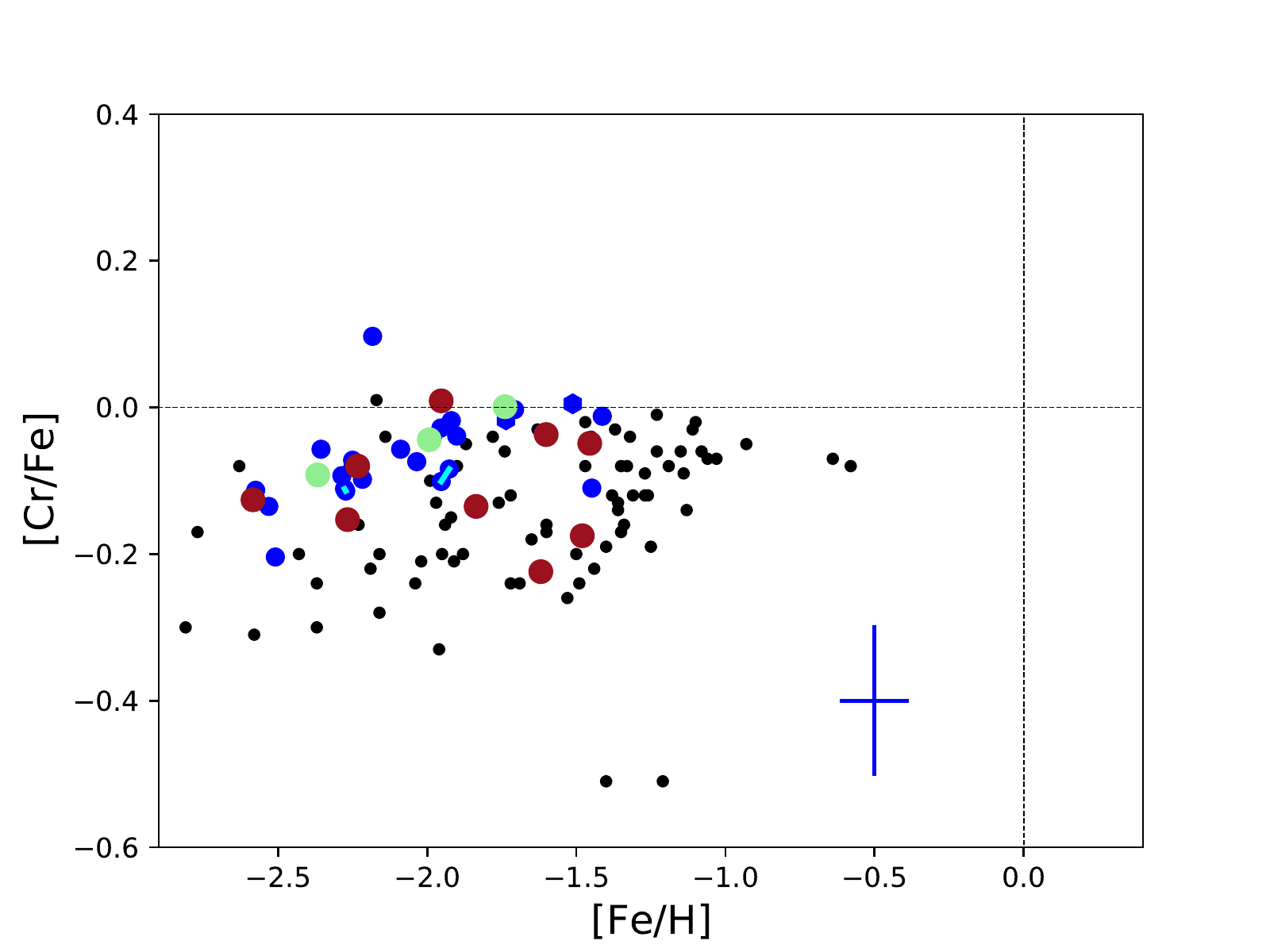}
\caption{[Cr/Fe] vs [Fe/H] abundances measured in the MINCE stars; the details are the same as Fig.\ref{OFe}, but without the lines of the models
and with the results
by \citet{ishigaki13} for halo stars (black dots).}
\label{CrFe}
\end{figure}

\begin{figure}[h]
\centering
\includegraphics[width=\hsize,clip=true]{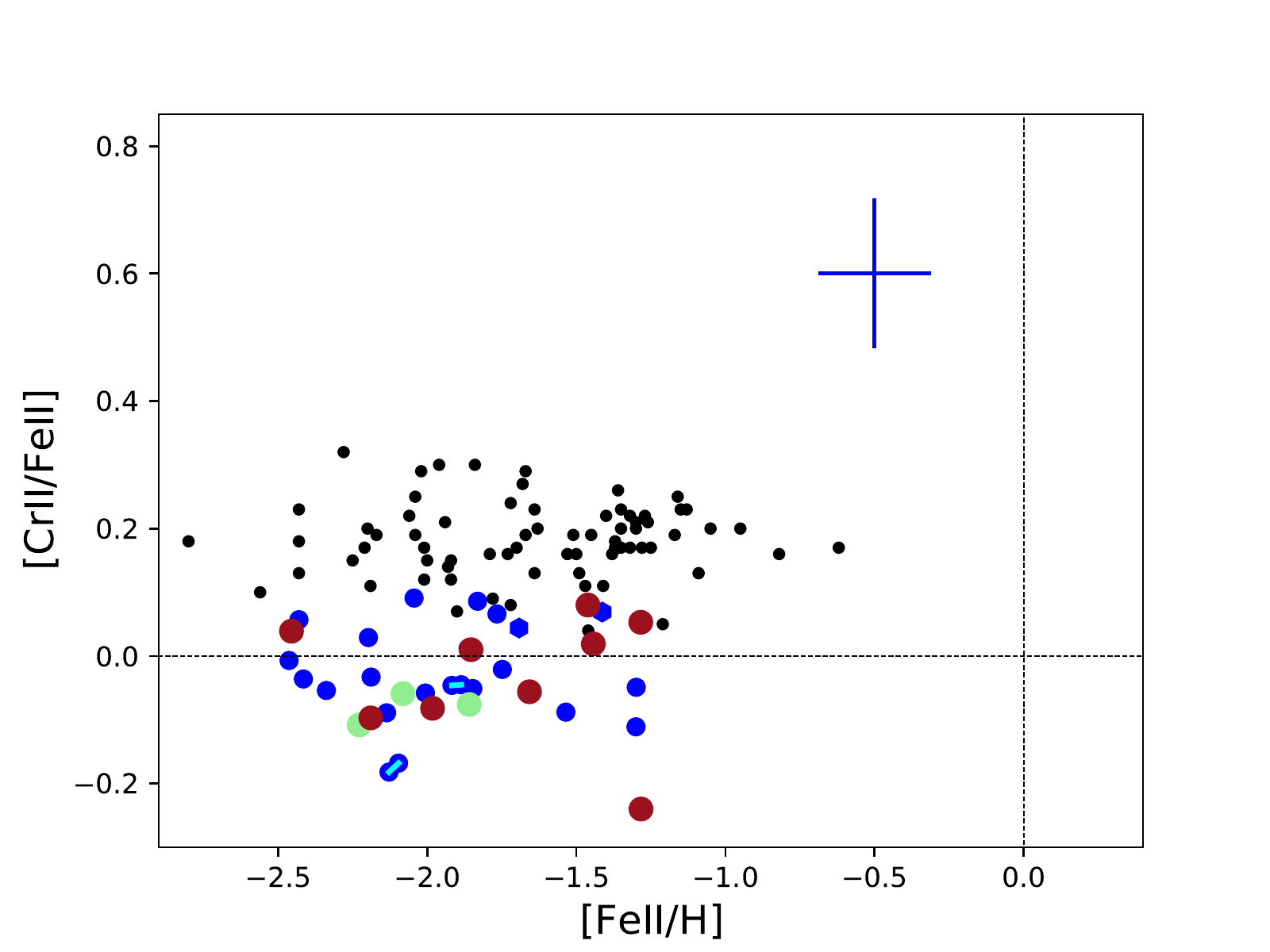}
\caption{[Cr II/Fe II] vs [Fe II/H] abundances measured in the MINCE stars; the details are the same as Fig.\ref{CrFe}.}
\label{CrIIFe}
\end{figure}

\begin{figure}[h]
\centering
\includegraphics[width=\hsize,clip=true]{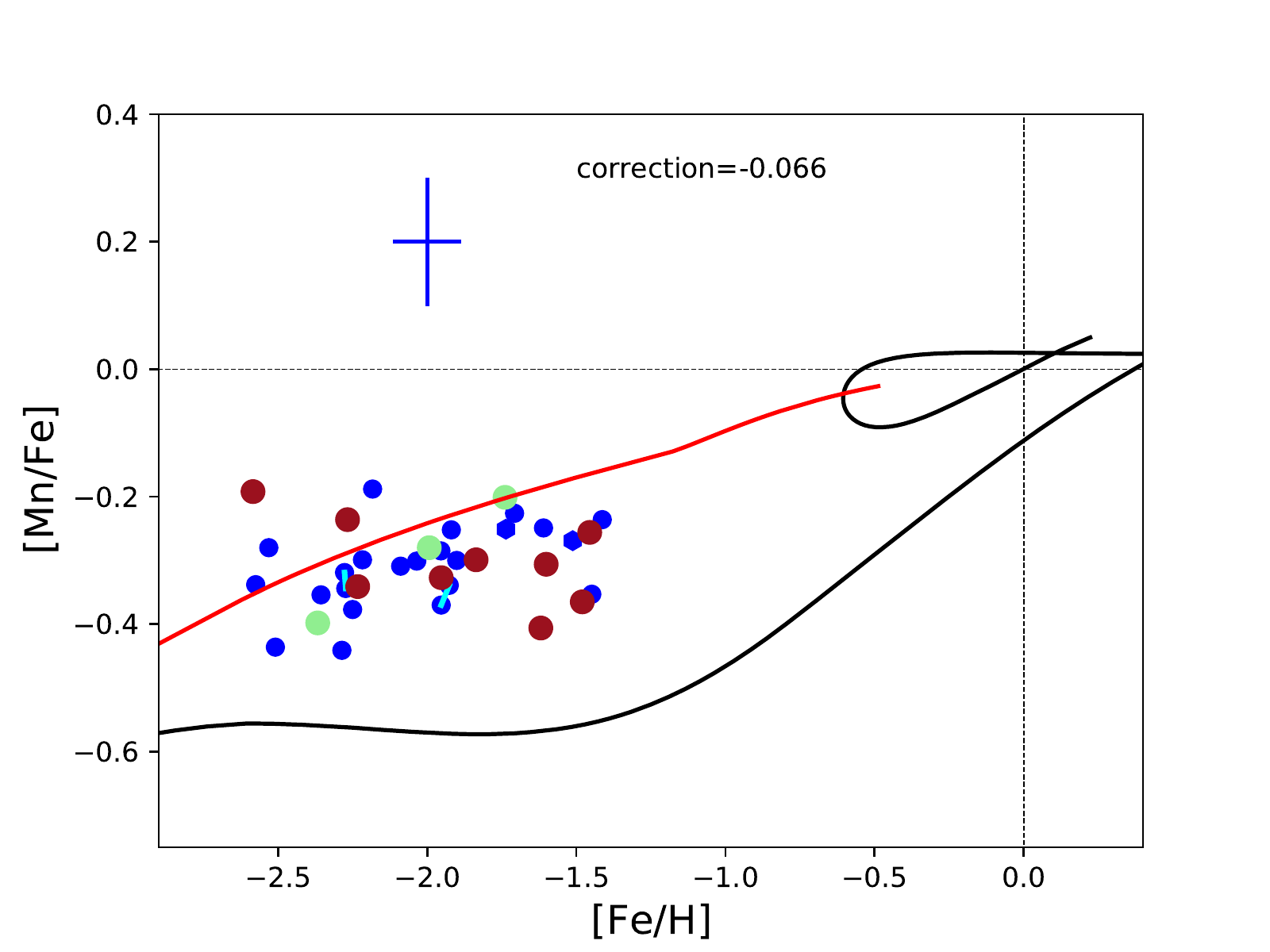}
\caption{[Mn/Fe] vs [Fe/H] abundances measured in the MINCE stars; the details are the same as Fig.\ref{OFe}.}
\label{MnFe}
\end{figure}

\begin{figure}[h]
\centering
\includegraphics[width=\hsize,clip=true]{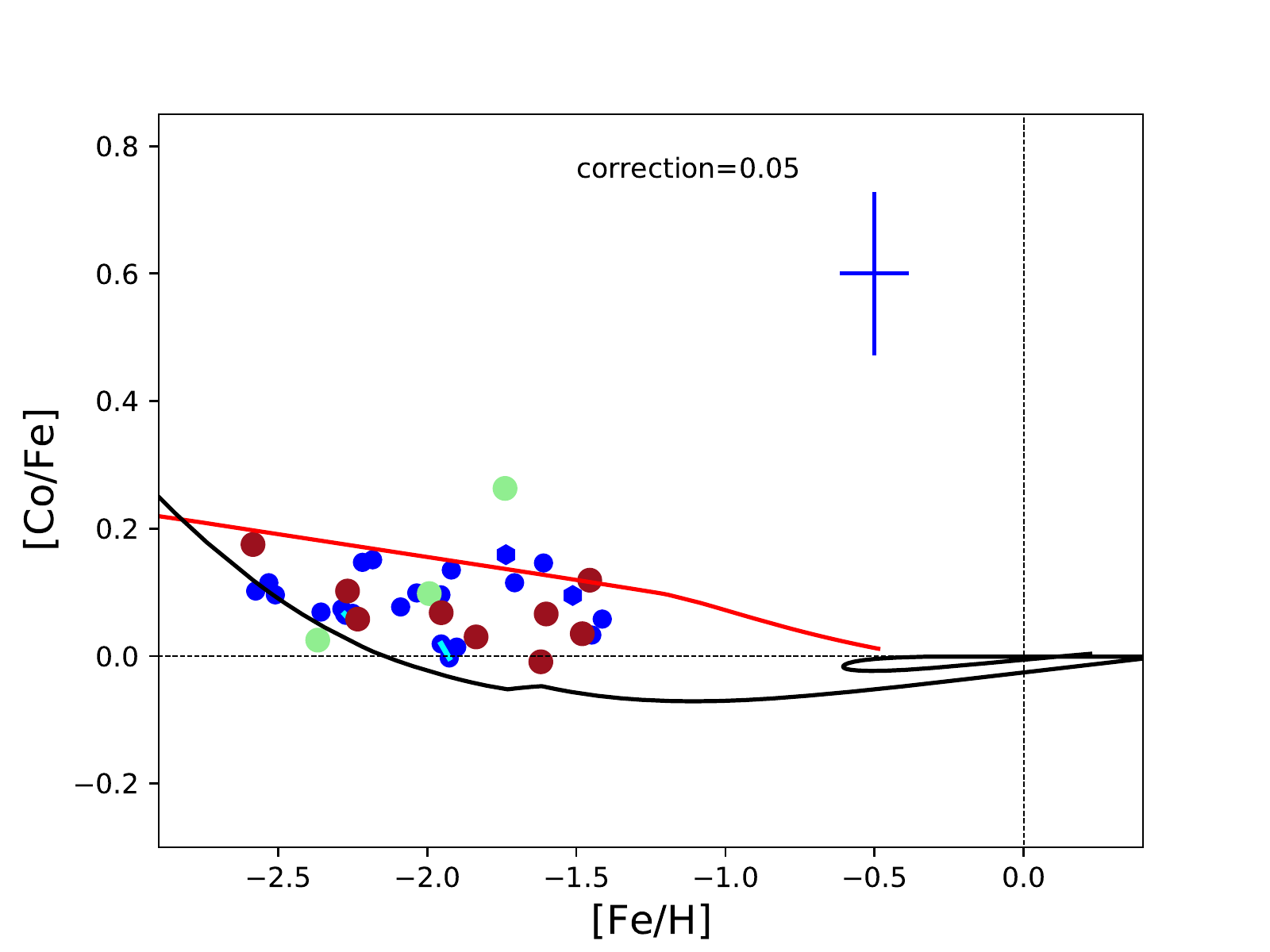}
\caption{[Co/Fe] vs [Fe/H] abundances measured in the MINCE stars; the details are the same as Fig.\ref{OFe}.
}
\label{CoFe}
\end{figure}

\begin{figure}[h]
\centering
\includegraphics[width=\hsize,clip=true]{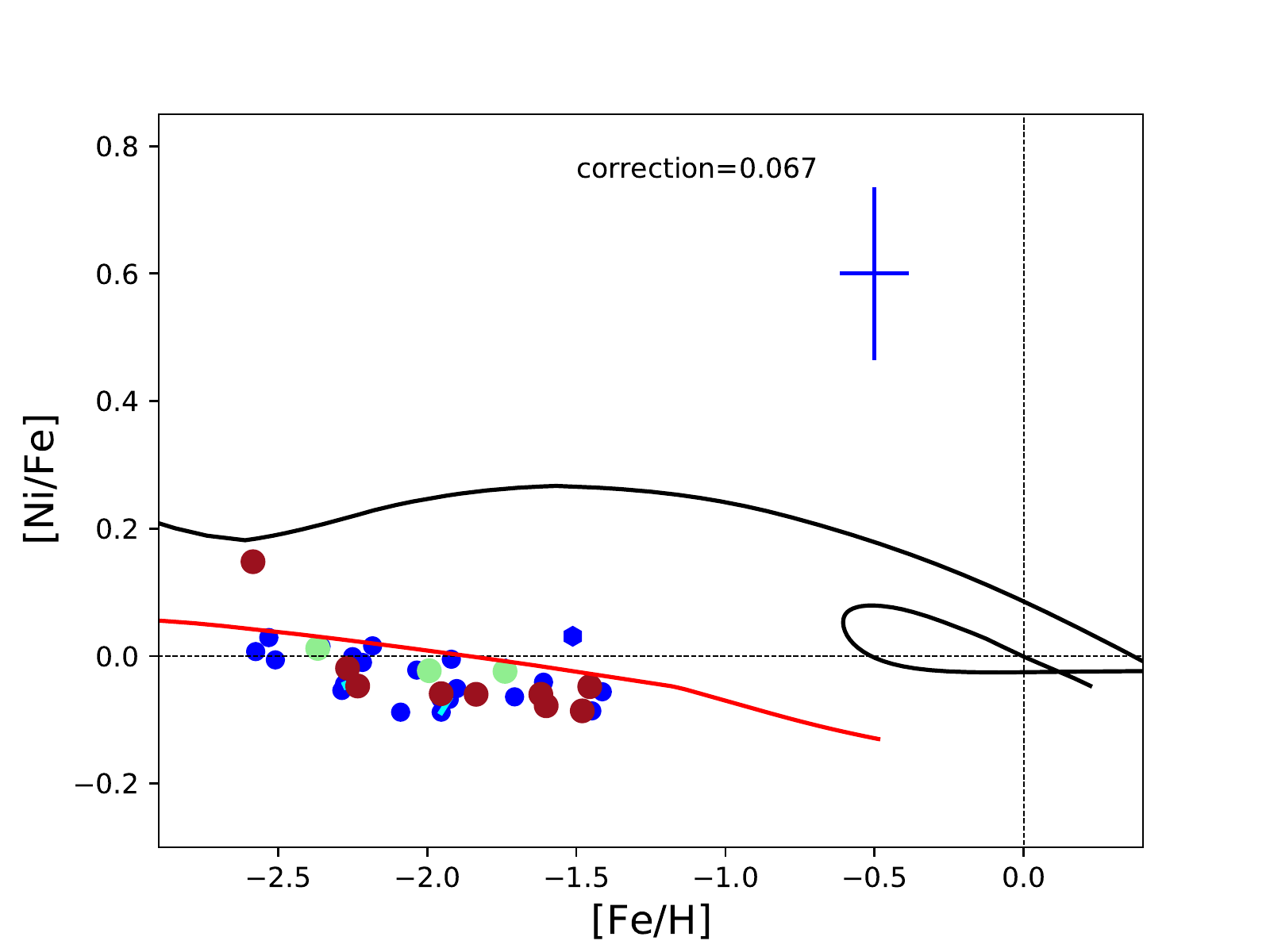}
\caption{[Ni/Fe] vs [Fe/H] abundances measured in the MINCE stars; the details are the same as Fig.\ref{OFe}.
}
\label{NiFe}
\end{figure}

\begin{figure}[h]
\centering
\includegraphics[width=\hsize,clip=true]{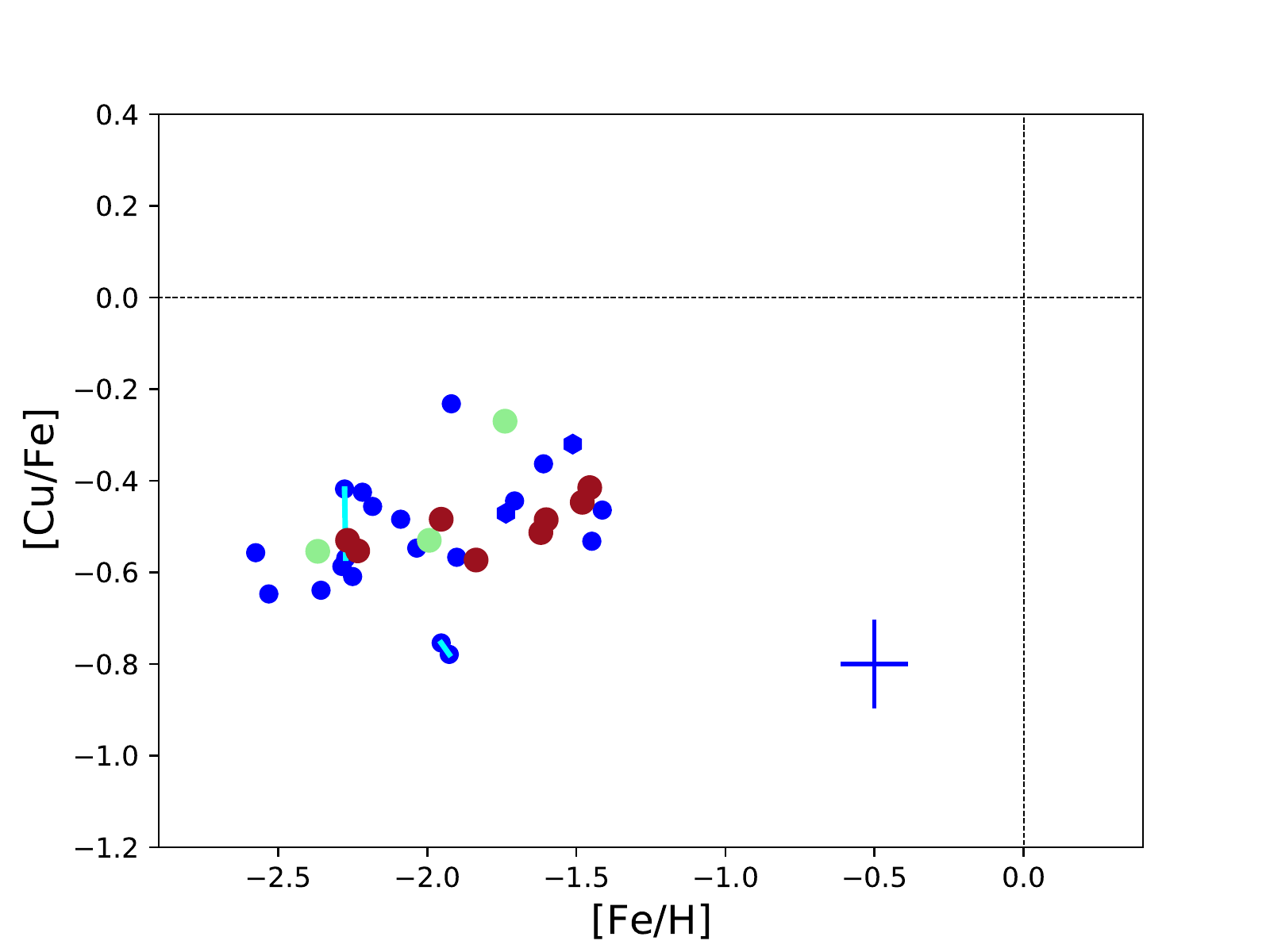}
\caption{[Cu/Fe] vs [Fe/H] abundances measured in the MINCE stars; the details are the same as Fig.\ref{OFe}, but without the lines of the models.}
\label{CuFe*}
\end{figure}

\begin{figure}[h]
\centering
\includegraphics[width=\hsize,clip=true]{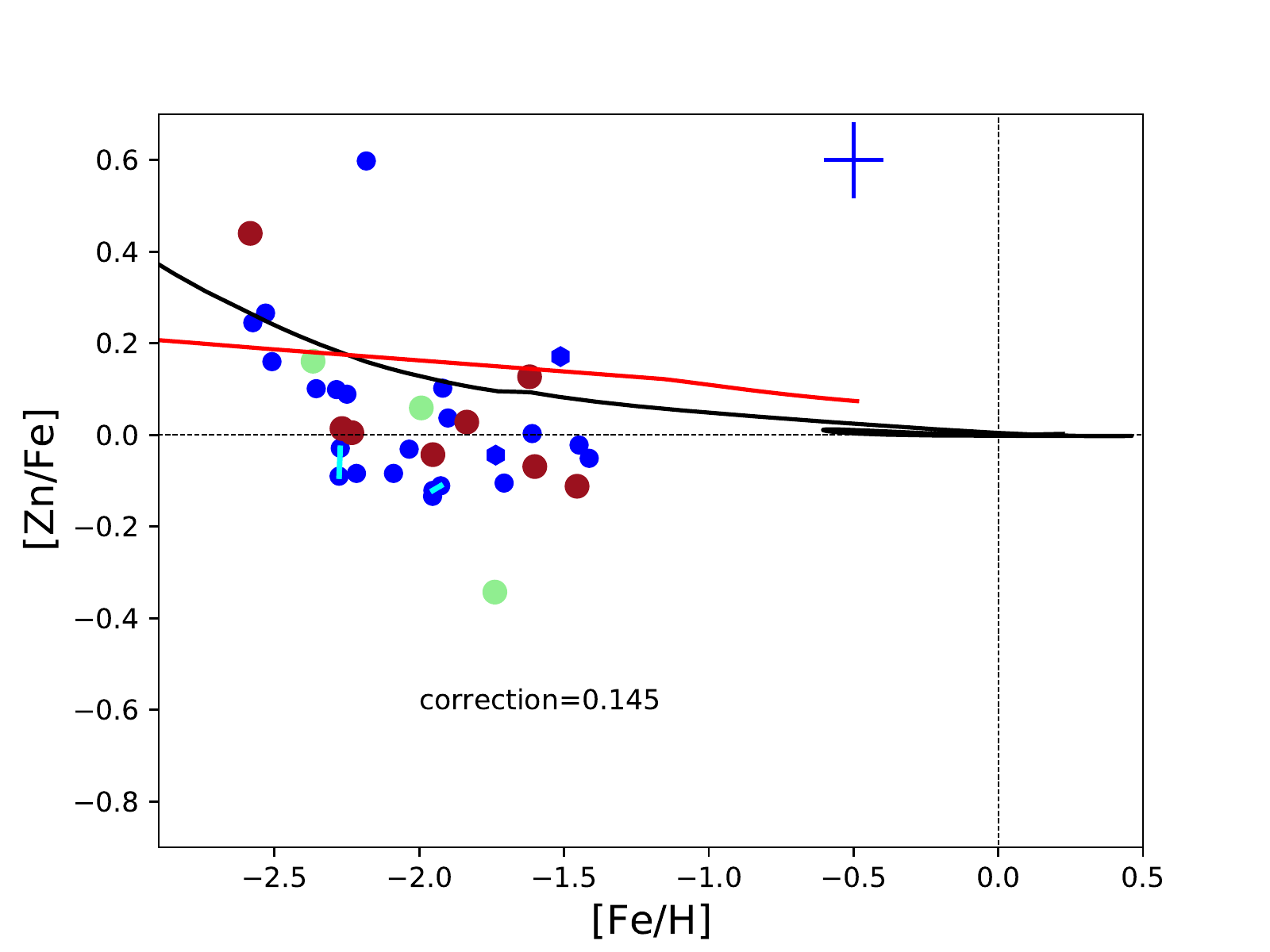}
\caption{[Zn/Fe] vs [Fe/H] abundances measured in the MINCE stars; the details are the same as Fig.\ref{OFe}.
}
\label{ZnFe}
\end{figure}

In Figs. \ref{ScFe}-\ref{ZnFe}, we present the results we obtained for the iron peak elements. Among them, copper presents the largest offset among the chemical abundance measurements of the two duplicate stars. The difference is anyway  $<$0.2 dex and for most of the other elements is well below 0.1 dex.  

Not surprisingly, most of the iron peak elements have a chemical evolution similar to the one of iron, being produced in a similar manner by SNe II and SNe Ia and, therefore, presenting a solar ratio. 
The most intriguing exceptions are manganese and copper that in the stellar atmosphere of our sample have negative abundance ratios compared to iron (normalised to the Sun). 
Manganese is a remarkable element, because it has a single stable isotope and for this reason its production is quite sensitive to the explosion conditions. In fact, theoretical computations have found that different classes of supernovae Ia are expected to produce  different amount of manganese \citep{Kobayashi15}.
Thanks to this characteristic, it was possible to exclude the exclusive enrichment of single degenerate SNe Ia  from chemical evolution modelling. It was also possible to evaluate the fraction of different SNe Ia contributing to the enrichment of manganese \citep{Seitenzahl13,Eitner20}, although the impact of NLTE in the determination and also the exact metallicity dependence of the yields of SNe II can impact the exact determination of this fraction. Moreover, the differential enrichment of manganese by the SNe Ia classes may also produce a spread in the enrichment, as shown in \citet{Cescutti17}.

On the other hand, copper is not expected to be significantly produced by SNe Ia, and the rise toward the solar metallicity is driven by a strong metal dependency in SNe II \citep{Timmes95}. 
Contrary to copper and manganese, scandium presents a behaviour similar to the one of the $\alpha$-elements, with a [Sc/Fe]$>$0 for [Fe/H]$<-$1.
This is controversial, in the sense that the results from \citet{Franc04} seem to indicate a  behaviour similar to standard iron peak elements, so approximately a constant [Sc/Fe]$\sim 0$. In this case, it is difficult also to rely to theoretical nucleosynthesis expectations, since the yields for scandium are usually too low by 1 dex \citep{Romano10,Kobayashi11}. 

The chemical evolution deduced from the MINCE stars for the rest of our iron peak elements appears to be remarkably similar to iron. We also note that our estimates for Cr I and Cr II are in agreement, contrary to the discrepancy observed in the \citet{ishigaki13} data for this element between ionised and neutral species; in fact, for this comparison data set it is present an average [Cr I/Fe I] ratio slightly below solar ratio, and slightly above for [Cr II/Fe II]. Also the different selection of Cr lines shall be accounted for  the discrepancy as also remarked in \citet{Lombardo}.
This trend of [Cr/Fe] is also compatible to the results obtained applying NLTE corrections for chromium in \citet{Bergemann10}.

Four stars appear to be enhanced in vanadium for [Fe/H]$<-$1.5 compared to rest of the sample.  Moreover, the stars with high [V/Fe] at [Fe/H]$\sim -$2.5  also show a high [Ni/Fe] and [Zn/Fe] as well as a low [Sc II/Fe II]; notably, this star belong the GSE substructure. 

We show the chemical evolution tracks also for iron peak elements, but we are afraid  that the yields assumed \citep[we recall that we use ][]{Franc04} are not the final answer, as shown already for manganese \citep{Cescutti08,Seitenzahl13,Cescutti17} and possibly true for other elements. Clearly, the chemical evolution models can only be as good as their nucleosynthesis input and the iron peak nucleosynthesis is not so well understood at present \citep[see e.g. Fig. 15 of][]{Kobayashi11}.
Still, the chemical evolution results for GSE seem to reproduce the observed ratios at least for manganese, cobalt, and nickel, indicating that the  role and timescale of SNe Ia are well considered. 

Again we do not see in the data any significant trend or offset between halo stars and GSE stars. On the other hand, the GSE star with the lowest iron content, BD+39\,3309 shows peculiar abundance ratios of [V/Fe], [Ni/Fe] and [Zn/Fe] that appear to be all enhanced with respect to sample. The interpretation of this enhancement is not trivial; [Zn/Fe] -- and to a lesser extent also [Ni/Fe] -- is expected to be higher in the nucleosynthesis of  hypernovae \citep{Kobayashi06}, compared to standard SNe II. Hypernovae belong to a class of SNe II exploding with a kinetic energy ten (or more) times the typical energy for SNe II of 10$^{51}$ erg and tend to eject a larger fraction of iron-peak elements. On the other hand, a hypernovae will be polluting with a lower ratios of [$\alpha$/Fe] and this does not seem to be the case. Certainly, we plan to monitor the presence of these stars -- enhanced in iron peak elements -- in future MINCE data. Regarding Sequoia stars, the abundance ratios of V, Cr, Mn, and Co as compared to Fe are increasing toward higher metallicity, whereas the opposite happens to Zn and Ni. Clearly with three stars, we cannot consider these trends to be significant, but again we will keep track of this hint in future MINCE papers.  

\section{NLTE corrections}\label{nlte}
Depending on the exact choice of lines combined with the stellar parameters and the abundances itself, some elemental abundances suffer from the 1D, LTE assumptions, while others remain good chemical tracers. Several studies have targeted such improvements by computing either NLTE or 3D abundances \citep[or both, see e.g.][]{Amarsi2019,Bergemann2017b,Bergemann2019,Caffau2008,Lind2012,Mashonkina2020,Sitnova2016,Steffen2015}.

The recent study by \cite{Hansen2020} presented corrected abundances for most of elements presented here, with the exception of Al. Owing to the overlap in stellar parameter space,
we used their NLTE computations as an indication of where corrections for the LTE assumptions would affect the LTE abundances most. A full abundance correction will be presented in a forth coming paper.
A few stars show a good agreement (overlap) in stellar parameters and the corrections, which are sensitive to the stellar parameters, can therefore help us assess the level or at least direction the NLTE corrections would
bring the corrected NLTE abundances in.
From \citet{Hansen2020} - BD-10\_3742 (T/logg/[Fe/H]/Vt: $4678\pm120/1.38\pm0.04/-1.96\pm0.07/1.9\pm0.1$) and BD-12\_106 ($4889\pm50/2.03\pm0.05/-2.11\pm0.04/1.5\pm0.2$) come close to two of the MINCE programme stars, namely BD+07\_4625 ($4757/1.64/-1.93/1.86$) and BD+39\_3309 ($4909/1.73/-2.58/1.94$). To estimate the order of magnitude of the corrections, we read off the NLTE corrections from their Table A.1, and we note that these are only approximate as the corrections also strongly depend on the use of lines and the actual size of the abundances as well. The NLTE corrections are presented as $\Delta$NLTE = NLTE $-$ LTE.

\begin{table}[!ht]
    \centering
    \caption{Approximate NLTE corrections for two MINCE stars.}
    \begin{tabular}{l c c}
\hline
H2020  & BD-10\_3742   &   BD-12\_106\\
MINCE  & BD+07\_4625   & BD+39\_3309 \\
\hline
Element & $\Delta NLTE$ & $\Delta NLTE$ \\
\hline
CI  & $-$0.06 &  $-$0.04 \\
OI  & $-$0.10 &  $-$0.10 \\
NaI &  $-$0.30 & $-$0.21 \\
MgI &  0.03  &  0.08 \\
SiI &  $-$0.13 & $-$0.12 \\
SI  &   ---  &  0.00 \\
KI  &  $-$0.16  &  $-$0.14 \\
CaI &   0.09  &  0.04 \\
ScII &  ---   & $-$0.03 \\
TiI  &  0.14  &  0.14 \\
TiII &  $-$0.03 &  $-$0.04 \\
CrI  &  0.16  &  0.21 \\
MnI  &  0.00  &  0.18 \\
FeI  &  0.08  &  0.11 \\
FeII &  0.00  &  0.02 \\
CoI  &  0.55  &  0.73 \\
\hline
    \end{tabular}
    \label{tab:NLTE}
\tablefoot{Based on \citet[][H2020]{Hansen2020}.}
\end{table}

From Table~\ref{tab:NLTE}, it is clearly seen that the largest corrections for such stars are obtained for Na, (SiI, K, TiI), Cr, Mn, and Co (especially the latter). In the case of Na, over recombination leads to strengthening of the lines and negative NLTE corrections.
For K the corrections also exceed $\pm0.1$dex, and here they are dictated by the source function, and caused by resonance line scattering, where similar to Na D lines an over population of the ground states shift the line formation outwards. This in turn deepens the K lines, so the effect is governed by the radiation field and rates. In this case, the values were interpolated using the grid from \citet{Reggiani2019}.
The corrections are positive for Si, where NLTE computations lead to weakened Ti I lines and, hence, result in positive corrections. Here the corrections are photoionisation dominated, which means that they are sensitive to overionisation driven by a non-local high-energy radiation field. This leads to weakening of low-excitation potential lines and in turn positive NLTE corrections (as LTE underestimates their abundances).
However, the largest corrections are seen for the Fe-peak elements, especially Co. For this element, the NLTE corrections may completely change the picture of its chemical evolution and surely the  nucleosynthesis adopted here, empirically deduced from LTE measurements, cannot be used, nor the original \citet{WW95} yields will improve the situation. Moreover, also other available tables of nucleosynthesis \citep{Kobayashi11,Limongi18} will struggle to explain the evolution of cobalt. Detailed NLTE studies for Mn and Co are important to properly understand their chemical evolution \citep[e.g.][]{Eitner20}. 

\section{Conclusions}
We describe the method adopted in the MINCE project to select  our sample, determine the stellar atmosphere of our stellar targets, and  measure at intermediate-low metallicity the chemical abundances of several $\alpha$-elements and iron peak elements, Na and Al.
The first selection criteria, based solely on Starhorse \citep{Anders19} was not ideal. It allowed us to properly select  the characteristics of the stars in term of $\logg$ and $\Teff$. It also correctly determines 
metal-poor stars, but not as metal-poor as requested by our project ([Fe/H]$<-1$). For this reason, we also implemented  a selection based on kinematics by  requiring the $v_{\rm tot} >$ 200\,km s$^{-1}$, so halo stars. 
With this new constraint, the selection is successful in finding stars with metallicities below [Fe/H]$<-$1 and therefore within the MINCE metallicity range.
Thanks to Gaia data, we were also able to distinguish among our sample those stars belonging to GSE (12) and Sequoia (3). We did not find specific trends and offsets compared to the sample of halo stars (defined as those not belonging neither to GSE nor Sequoia). This is not completely unexpected given that the sample is still small; moreover, the chemical evolution results also did not predict the important feature in the metallicity range that we explore here -- but it did indeed do so for slightly more metal-rich objects. 
The results of this first campaign show that the approach of using multiple middle-sized facilities allows to 
collect meaningful amounts of high-quality data in a short time. In the next paper of the series, we shall
present the measurements of neutron capture elements in this sample of stars. 

\begin{acknowledgements}
We gratefully acknowledge support from the French National Research Agency (ANR) funded project ``Pristine'' (ANR-18-CE31-0017). This work was partially supported by the European Union (ChETEC-INFRA, project no. 101008324).
This work has made use of data from the European Space Agency (ESA) mission
{\it Gaia} (\url{https://www.cosmos.esa.int/gaia}), processed by the {\it Gaia}
Data Processing and Analysis Consortium (DPAC,
\url{https://www.cosmos.esa.int/web/gaia/dpac/consortium}). Funding for the DPAC
has been provided by national institutions, in particular the institutions
participating in the {\it Gaia} Multilateral Agreement.
This research has made use of the SIMBAD database, operated at CDS, Strasbourg, France. ES received funding from the European Union’s Horizon 2020 research and innovation program under SPACE-H2020 grant agreement number 101004214 (EXPLORE project). Funding for the Stellar Astrophysics Centre is provided by The Danish National Research Foundation (Grant agreement no.: DNRF106).
\end{acknowledgements}

\bibliographystyle{aa}
\bibliography{spectro}

\begin{appendix}
\onecolumn

\section{Log of the observations}

\begin{table}[h]
\caption{\label{harpsn} Log of the observations and radial velocities for the stars observed with HARPS-N}
\begin{tabular}{lrrrrrrr}
\hline
  \multicolumn{1}{c}{Star} &
  \multicolumn{1}{c}{alpha2000} &
  \multicolumn{1}{c}{delta2000}  &
\multicolumn{1}{c}{BJD} &
\multicolumn{1}{c}{Date} &
\multicolumn{1}{c}{$t_{exp}$} &
\multicolumn{1}{c}{RV}&
\multicolumn{1}{c}{$\sigma_{RV}$}
\\
  \multicolumn{1}{c}{~   } &
  \multicolumn{1}{c}{J2000} &
  \multicolumn{1}{c}{J2000} &
\multicolumn{1}{c}{days} &
\multicolumn{1}{c}{} &
\multicolumn{1}{c}{$s$} &
\multicolumn{1}{c}{$\rm km\,s^{-1} $}&
\multicolumn{1}{c}{$\rm km\,s^{-1} $}
\\

\\
\hline
  HD\,87740      & 10:07:10.25 &  +03:41:23.3 & 2458980.40412383 & 2020-05-10 & 3000 & --23.9993 & 0.0016\\
  HD\,91276      & 10:32:57.37 &  +35:22:56.6 & 2458980.44256365 & 2020-05-10 & 3000 &  +23.9350 & 0.0017 \\
  BD\,+13\,2383  & 11:17:37.07 &  +12:24:10.0 & 2458980.48078609 & 2020-05-10 & 2400 & --11.5311 & 0.0023 \\
  BD\,+41\,2520  & 14:42:02.54 &  +41:14:11.6 & 2458980.60643665 & 2020-05-11 & 3600 &  +11.6820 & 0.0014 \\
  HD\,130971     & 14:51:15.68 & --08:59:01.8 & 2458980.56913504 & 2020-05-11 & 3600 &  +23.7756 & 0.0012\\
  BD\,+24\,2817  & 15:05:56.81 &  +24:05:51.7 & 2458980.65155434 & 2020-05-11 & 3000 & --43.1845 & 0.0009 \\
  HD\,138934     & 15:34:21.37 &  +23:12:36.6 & 2458980.68241001 & 2020-05-11 & 2100 &  +18.3412 & 0.0007\\
  HD\,143348     & 15:58:36.55 &  +34:11:33.4 & 2458980.70923832 & 2020-05-11 & 2400 & --73.7397 & 0.0012\\
  BD\,--07\,3523 & 13:00:33.60 & --07:59:38.2 & 2459010.45002716 & 2020-06-09 & 3600 &  +73.5710 & 0.0025\\
  HD\,115575     & 13:18:09.97 & --13:58:45.8 & 2459010.40738054 & 2020-06-09 & 3600 & +188.4806 & 0.0029\\
  BD\,+06\,2880  & 14:25:10.31 &  +06:07:14.9 & 2459010.49153284 & 2020-06-09 & 3000 &  +37.3641 & 0.0021\\
  HD\,238439     & 15:17:00.58 &  +54:35:38.6 & 2459010.53202863 & 2020-06-10 & 3600 & --65.0949 & 0.0024\\
  HD\,139423     & 15:37:45.83 &  +11:36:11.6 & 2459010.56447555 & 2020-06-10 & 1200 & +183.3917 & 0.0023 \\
  HD\,142614     & 15:55:15.38 &  +08:13:27.8 & 2459010.58578868 & 2020-06-10 & 2100 & -337.1939 & 0.0019\\
  BD\,+254520    & 21:22:08.32 &  +25:45:15.8 & 2459010.61970877 & 2020-06-10 & 3600 & +22.8268  & 0.0025 \\
  HD\,208316     & 21:55:36.03 & --04:13:27.4 & 2459010.65672317 & 2020-06-10 & 2100 & --146.1986& 0.0019\\
\hline\end{tabular}
\end{table}

\begin{table}[h]
\caption{\label{fies} Log of the observations and radial velocities for the stars observed with FIES}
\begin{tabular}{llllllrr}
\hline
  \multicolumn{1}{c}{STAR} &
  \multicolumn{1}{c}{$\alpha$} &
  \multicolumn{1}{c}{$\delta$} &
\multicolumn{1}{c}{BJD} &
\multicolumn{1}{c}{Date} &
\multicolumn{1}{c}{$t_{exp}$} &
\multicolumn{1}{c}{RV}&
\multicolumn{1}{c}{$\sigma_{RV}$}
\\
  \multicolumn{1}{c}{~   } &
  \multicolumn{1}{c}{J2000} &
  \multicolumn{1}{c}{J2000} &
\multicolumn{1}{c}{days} &
\multicolumn{1}{c}{} &
\multicolumn{1}{c}{$s$} &
\multicolumn{1}{c}{$\rm km\,s^{-1} $}&
\multicolumn{1}{c}{$\rm km\,s^{-1} $}
\\
\hline
  BD\,+07\,4625  & 21:07:13.10 & +07:44:19.8 & 2459032.670574271 &2020-07-02& 2200 & --494.883 & 0.004\\
                 & 21:07:13.10 & +07:44:19.8 & 2459032.696585007 &2020-07-02& 2200 & --494.551 & 0.003\\
  BD\,+35\,4847  & 22:37:13.45 & +36:08:21.6 & 2459033.675984722 &2020-07-03& 2800 & --139.739 & 0.003\\
                 & 22:37:13.45 & +36:08:21.6 & 2459033.708940891 &2020-07-03& 2800 & --139.742 & 0.003\\
  BD\,+11\,2896  & 16:01:04.87 & +11:12:56.2 & 2459001.539097893 &2020-06-01& 3000 & --218.821 & 0.002\\
                 & 16:01:04.87 & +11:12:56.2 & 2459001.574364527 &2020-06-01& 3000 & --218.831 & 0.002\\
  HD\,165400     & 18:05:30.45 & +09:49:30.4 & 2459000.690305141 &2020-05-31& 2800 &    --2.745& 0.002\\
  BD\,--00\,3963 & 20:17:12.53 & +00:21:22.7 & 2459036.622693848 &2020-07-06& 2200 &   --42.165& 0.002\\
                 & 20:17:12.53 & +00:21:22.7 & 2459036.648704157 &2020-07-06& 2200 &   --42.159& 0.002\\
  BD\,--00\,4538 & 23:38:18.78 & +00:46:51.5 & 2459069.603750275 &2020-08-08& 2900 &  --190.753& 0.003 \\
                 & 23:38:18.78 & +00:46:51.5 & 2459069.637862387 &2020-08-08& 2900 &  --191.027& 0.003\\
\hline\end{tabular}
\end{table}

\begin{table}
\caption{\label{ohp} Log of the observations and radial velocities for the stars observed with Sophie}
\begin{tabular}{llllllrr}
\hline
  \multicolumn{1}{c}{STAR} &
  \multicolumn{1}{c}{$\alpha$} &
  \multicolumn{1}{c}{$\delta$} &
\multicolumn{1}{c}{BJD} &
\multicolumn{1}{c}{Date} &
\multicolumn{1}{c}{$t_{exp}$} &
\multicolumn{1}{c}{RV}&
\multicolumn{1}{c}{$\sigma_{RV}$}
\\
  \multicolumn{1}{c}{~   } &
  \multicolumn{1}{c}{J2000} &
  \multicolumn{1}{c}{J2000} &
\multicolumn{1}{c}{days} &
\multicolumn{1}{c}{} &
\multicolumn{1}{c}{$s$} &
\multicolumn{1}{c}{$\rm km\,s^{-1} $}&
\multicolumn{1}{c}{$\rm km\,s^{-1} $}
\\
\hline
  TYC\,4-369-1 & 00:08:36.02 & +02:58:01.7& 2459087.56089588& 2020-08-25 & 3600 & +3.982 & 0.006 \\
  BD\,+04\,18 & 00:12:49.90 & +05:37:39.3 & 2459086.6077017& 2020-08-25 & 3600 & --29.987 & 0.002\\
  TYC\,33-446-1 & 01:54:22.17 & +03:41:45.3 & 2459087.64126008 & 2020-08-26 & 3491 & --99.892 & 0.004 \\
  TYC\,2824-1963-1 & 01:58:38.93 & +41:46:30.4 & 2459086.6437657 & 2020-08-25 & 2618 & +52.907 & 0.003\\
                   & 01:58:38.93 & +41:46:30.4 & 2459087.5992952 & 2020-08-26 & 3105 & +52.816 & 0.002\\
  TYC\,4331-136-1 & 03:57:14.19 & +69:44:45.1  & 2459088.63728359& 2020-08-27 & 3600 & --110.967 & 0.005\\
  TYC\,1008-1200-1 & 18:06:31.58 & +08:44:54.7 & 2459086.3381710 & 2020-08-24 & 3600 & --393.813 & 0.007\\
                   & 18:06:31.58 & +08:44:54.7 & 2459087.3363853 & 2020-08-25 & 3600 & --393.838 & 0.005\\
  TYC\,2113-471-1 & 18:56:41.55 & +25:16:50.8  & 2459087.38188494& 2020-08-25 & 3600 & --252.760 & 0.004\\
                  & 18:56:41.55 & +25:16:50.8  & 2459088.32588031& 2020-08-26 & 3600 & --252.858 & 0.004\\
  TYC\,4221-640-1 & 19:09:19.27 & +63:03:44.2 & 2459086.38173932& 2020-08-24 & 3600 &       &      \\
                  & 19:09:19.27 & +63:03:44.2 & 2459088.36663955& 2020-08-26 & 3600 &--277.441& 0.006 \\
  TYC\,4584-784-1 & 19:22:56.40 & +76:32:43.3 & 2459088.41633999& 2020-08-26 & 3600 & --295.607&0.004\\
  TYC\,3944-698-1 & 20:02:59.61 & +58:01:07.1 & 2459086.4263490 &2020-08-24 & 3600 &  --255.266 &0.004\\
  HD\,354750 & 20:04:29.05 & +13:35:31.0      & 2459088.46611671&2020-08-26 & 3600 &  --168.281 &0.008\\
  BD\,+25\,4520 & 21:22:08.32 & +25:45:15.8   & 2459087.42848911&2020-08-25 & 3600 & +23.518     &0.003 \\
  TYC\,4267-2023-1 & 22:01:46.08 & +62:27:40.6& 2459086.4710943 & 2020-08-24 & 3600 & --346.268 & 0.003\\
  TYC\,565-1564-1 &22:10:38.77 &+05:16:14.6 & 2459087.47351081& 2020-08-25 & 3600 & --175.181 & 0.003\\
  BD\,+21\,4759 & 22:28:46.35 & +22:09:11.4     & 2459088.59697759& 2020-08-27 & 3600 & --202.045 & 0.006\\
  TYC\,2228-838-1 & 22:38:23.28 & +27:34:24.7 & 2459088.5536172 & 2020-08-27 & 3600 & --145.006 & 0.002\\
  TYC\,4001-1161-1 & 23:47:30.68 & +53:47:16.5 & 2459086.5158160 & 2020-08-24 & 3600 & -397.649 & 0.003 \\
                   & 23:47:30.68 & +53:47:16.5 & 2459087.5143166 & 2020-08-25 & 3506 & -397.617 & 0.003 \\
  BD\,+03\,4904 & 23:55:28.37 & +04:21:17.9& 2459086.5634472 & 2020-08-25 & 3600 & --208.370 & 0.006\\
  BD\,+07\,4625 & 21:07:13.10 & +07:44:19.7& 2459088.51050063 &2020-08-27& 3600 & --495.699 & 0.002\\
\hline\end{tabular}
\end{table}

\begin{table}
\caption{\label{espadons} Log of the observations and radial velocities for the stars observed with ESPaDOnS}
\begin{tabular}{llllllrr}
\hline
  \multicolumn{1}{c}{STAR} &
  \multicolumn{1}{c}{$\alpha$} &
  \multicolumn{1}{c}{$\delta$} &
\multicolumn{1}{c}{BJD} &
\multicolumn{1}{c}{Date} &
\multicolumn{1}{c}{$t_{exp}$} &
\multicolumn{1}{c}{RV}&
\multicolumn{1}{c}{$\sigma_{RV}$}
\\
  \multicolumn{1}{c}{~   } &
  \multicolumn{1}{c}{J2000} &
  \multicolumn{1}{c}{J2000} &
\multicolumn{1}{c}{days} &
\multicolumn{1}{c}{} &
\multicolumn{1}{c}{$s$} &
\multicolumn{1}{c}{$\rm km\,s^{-1} $}&
\multicolumn{1}{c}{$\rm km\,s^{-1} $}
\\
\hline
  BD+20  3298 & 16:36:33.15 & +20:25:46.1     & 2459016.033335 & 2020-06-15 & 2380 & --257.104& 0.003 \\
  BD+31  2143 & 10:28:17.23 & +30:26:29.2     & 2459180.153748 & 2020-11-26 & 2380 &  +64.157& 0.004  \\
  BD+48  2167 & 13:59:19.74 & +48:05:35.5     & 2459189.153872 & 2020-12-05 & 2380 & --108.203& 0.003  \\
  BD+39  3309 & 18:03:47.35 & +39:32:31.3     & 2459016.091233 & 2020-06-15 & 2380 & --249.092& 0.005  \\
  BD+32  2483 & 14:31:38.96 & +31:58:58.4     & 2459012.887960 & 2020-06-12 & 2380 &    +4.145 & 0.003  \\
  TYC 3085-119-1 & 17:16:36.98 & +44:10:43.4  & 2458739.761057 & 2019-09-13 & 2380 & --106.234& 0.003  \\
                 & 17:16:36.98 & +44:10:43.4  & 2458739.790407 & 2019-09-13 & 2380 & --105.941& 0.003  \\
  TYC 2588-1386-1 & 16:41:32.08 & +36:24:42.6 & 2458739.725085 & 2019-09-13 & 2380 & --249.674& 0.003 \\
\hline\end{tabular}
\end{table}
\newpage
\section{Linelist}

\begin{table}[h]\caption{Example of the table available at the CDS with the list of the atomic data of the lines measured for each star of the MINCE sample}\label{linelist}
 \begin{tabular}{|l|r|r|l|r|r|r|}
\hline
  \multicolumn{1}{|c|}{Element} &
  \multicolumn{1}{c|}{Z} &
  \multicolumn{1}{c|}{ion} &
  \multicolumn{1}{c|}{Star} &
  \multicolumn{1}{c|}{Wavelength} &
  \multicolumn{1}{c|}{Loggf} &
  \multicolumn{1}{c|}{Elow} \\
  &&&&nm\,\,&cm$^{-1}$&\\
\hline
  O & 8 & 0 & BD+03 4904 & 636.3776 & -10.19 & 161.311\\
  Mg & 12 & 0 & BD+03 4904 & 457.1096 & -5.623 & 0.0\\
  Mg & 12 & 0 & BD+03 4904 & 470.2991 & -0.44 & 35052.859\\
  Mg & 12 & 0 & BD+03 4904 & 552.8405 & -0.498 & 35052.859\\
  Mg & 12 & 0 & BD+03 4904 & 571.1088 & -1.724 & 35052.859\\
  Si & 14 & 0 & BD+03 4904 & 568.4484 & -1.553 & 39956.711\\
  Ca & 20 & 0 & BD+03 4904 & 452.6928 & -0.548 & 21849.562\\
  Ca & 20 & 0 & BD+03 4904 & 457.8551 & -0.697 & 20333.238\\
  ...&...&...&...&...&...&...\\
  
\end{tabular}
\end{table}

\begin{landscape}

\section{Chemical abundances}
\begin{table}[h!]
\setlength\tabcolsep{3pt}
\caption{Example of the first part of table available at the CDS with the abundances of elements in their ionisation state from O I to Sc II in [X/H].} 

\label{table_abbA}
  \begin{tabular}{llrccccccccccccccccccccc}
  \hline
Star & Spect. &S/N@550\,nm &\ion{O}{I} &         $\sigma$ &      N &     \ion{Na}{I} &    $\sigma$ &      N  &    \ion{Mg}{I} &   $\sigma$ &      N  &     \ion{Al}{I} &   $\sigma$ &      N &     \ion{Si}{I} &   $\sigma$ &        N  &    \ion{Ca}{I} &   $\sigma$ &      N  &    \ion{Sc}{II} &  $\sigma$ &       N\\
\hline
BD-00 4538     & FIES  & $> 100$               &  -1.12 & 0.0 & 1 & -2.11 & 0.05 & 3 & -1.42 & 0.11 & 4 & -99.0 & 0.0 & 0 & -1.52 & 0.05 & 14 & -1.55 & 0.06 & 25 & -1.54 & 0.12 & 10 \\
...
\end{tabular}
\tablefoot{Sigma is based on the line to line dispersion and N is the number of lines for each ion.}
\end{table}
\begin{table}[h!]
\setlength\tabcolsep{3pt}
\caption{Second part of the table available at the CDS with the abundances of elements from Ti I to Mn I in [X/H].}\label{table_abbB}
\begin{tabular}{lrccccccccccccccccccccc}
\hline
  Star & Ti I & $\sigma$ & N & Ti II & $\sigma$ & N  & V I & $\sigma$ & N & Cr I & $\sigma$ & N  & Cr II & $\sigma$ & N  & Mn I & $\sigma$  & N \\
 \hline
BD-00 4538 & -1.63 & 0.07 & 52 & -1.47 & 0.12 & 28 & -1.96 & 0.10 & 20 & -1.93 & 0.09 & 23 & -1.77 & 0.10 & 7 & -2.20 & 0.07 & 14 \\
...
\end{tabular}
\end{table}

\begin{table}[h!]
\setlength\tabcolsep{3pt}
\caption{Third part of the table available at the CDS with the abundances of elements from Fe I to Zn I in [X/H].}\label{table_abbC}
\begin{tabular}{lrccccccccccccccccccccc}
\hline
Star & Fe I & $\sigma$ & N & Fe II & $\sigma$ & N & Co I & $\sigma$ & N  & Ni I & $\sigma$ & N & Cu I & $\sigma$  & N  & Zn I & $\sigma$ & N \\
\hline
BD-00 4538 &  -1.90 & 0.09 & 276 & -1.75 & 0.14 & 26 & -1.89 & 0.09 & 19 & -1.95 & 0.10 & 64 & -2.47 & 0.15 & 4 & -1.87 & 0.03 & 2 \\
...
\end{tabular}
\end{table}

\end{landscape}

\end{appendix}






%




%



\end{document}